\documentclass[10pt,a4paper]{scrartcl}
\usepackage[utf8]{inputenc}
\usepackage[shortlabels]{enumitem}
\setkomafont{disposition}{\normalcolor\bfseries}
\usepackage[affil-it]{authblk}

\usepackage{amsmath}
\usepackage{appendix}
\usepackage{enumitem}
\usepackage{adjustbox}
\usepackage{fancybox}
\usepackage{framed}
\usepackage{setspace}
\usepackage{todonotes}
\usepackage{csquotes}
\usepackage{tikz}
\usepackage{eurosym}
\usepackage{epstopdf}
\usepackage{tabularx}
\usepackage{threeparttable}
\usepackage{booktabs}
\usepackage{multirow}
\usepackage{ragged2e}
\usepackage{csquotes}
\usepackage{float}
\usepackage{placeins}
\usepackage[colorlinks=true, allcolors=blue, breaklinks=true,hyperindex,breaklinks]{hyperref}
\DeclareGraphicsExtensions{.eps}
\usepackage{natbib}
\bibpunct[, ]{(}{)}{,}{a}{}{,}%
\def\bibfont{\small}%

\newtheorem{hypothesis}{Hypothesis}

\usepackage{multirow}
\usepackage{multicol}
\usepackage{subcaption}
\usepackage{csquotes}
\usepackage{comment}

\begin{document}
%\RUNTITLE{Utility Model}

%\TITLE{UM as a learning mechanism to produce high impact and specialized frontier knowledge}
%\TITLE{The utility model as a learning mechanism: from incremental adaptation to specialized frontier innovation}
% Alternative titles: 
%\TITLE{Becoming a global innovation leader: the utility model as a learning device}
%\TITLE{From imitation to global innovation leadership: the utility model as a learning device} 
%\TITLE{From imitation to creativity: The utility model as a learning device bridging from a catch-up to post-catch-up economy\\[1em]} 
%\TITLE{From imitation to creativity: The utility model as a learning device bridging the catch-up and post-catch-up economies\\[1em]} 
%\TITLE{From imitation to global frontier innovation: The utility model as a learning device to escape the middle-income trap\\[1em]} 
\title{From catch-up to frontier: The utility model \\as a learning device to escape the middle-income trap\\[1em]} 
%\TITLE{From imitator to global innovation leader: the utility model as a learning device in the Korean catch-up}
%\TITLE{From imitator to global innovation leader: the utility model as a learning device in Korea}
%\TITLE{From imitator to global innovation leader: the utility model as a learning device in technological catch-up}

\author[1,2]{\normalsize Su Jung Jee\thanks{Contacts: s.j.jee@sheffield.ac.uk (Corresponding Author); kerstin.hotte@kedgebs.com
}}
\author[3]{\normalsize Kerstin H\"otte}
\affil[1]{\normalsize Sheffield University Management School, University of Sheffield, UK}
\affil[2]{\normalsize Institute for New Economic Thinking, University of Oxford, UK}
\affil[3]{\normalsize Kedge Business School, Paris Campus, FR}
\date{\normalsize\today}
\maketitle

\begin{abstract}
     %\textcolor{LB}{Escaping} the middle-income trap requires a country to develop indigenous technological capabilities for high value-added innovation.
The second-tier patent system, known as utility models (UMs), has been considered a tool for supporting technological learning in catching-up economies. UMs often impose lower inventive step requirements than patents, making them accessible to inventors with limited capacity. However, the mechanisms by which the UM system supports the catch-up process and the system's long-term impact in the post-catch-up period remain underexplored. We address this gap by empirically examining how and under which conditions the UM system can create synergies with catch-up strategies and influence the emerging industrial structure in the post-catch-up phase. Drawing on the literature on catch-up and intellectual property rights (IPR), we argue that catch-up strategies that prioritize imitative and adaptive learning in short-cycle technologies can be complemented by the UM system, supporting the development of indigenous capabilities to produce high-impact frontier technologies and internalize their value-added. Using South Korea as a case study, we present two key findings: First, the country's post-catch-up frontier technologies (US patents) are more impactful when they build on active learning experiences encoded in domestic UMs. Second, UM-reliant frontier technologies contribute to value internalization (local value-added relative to foreign value-added). Our findings extend the literature on the role of different types of IPR in various stages of development and inform policy discussions on how UMs can be used to support capability building in developing countries.
%we theorize how a catch-up strategy that prioritizes imitative and adaptive learning in short-cycle technologies can be complemented by the UM system, and how this, in turn, supports the development of indigenous capabilities to produce high-impact frontier technologies and internalize their value-added. 
\end{abstract}

\vspace{0.2cm}
\textbf{Keywords:} utility model, intellectual property rights, technological catch-up, specialization, localization, post-catch-up

\newpage

\section*{Acknowledgements}
This article benefitted from extensive comments by Jussi Heikkila. The authors thankfully acknowledge helpful feedback received by staff members of the Gesellschaft f\"ur Internationale Zusammenarbeit (GIZ) and academic peers from the INET complexity group. 

\section*{Highlights}
\begin{itemize}
\item Evidence of the UM system’s supporting role in indigenous capability building in a catching-up economy and its long-term impact
\item The UM system can create synergy with a strategy that prioritizes imitative and adaptive learning in short-cycle technologies.
\item UM-based learning can lay the foundation for producing impactful frontier technologies.
\item UM-based learning helps internalize the value-added generated by these frontier technologies.
\end{itemize}
\color{black}
\newpage

\section{Introduction} 
Technological catch-up is essential for economic development as countries move from import dependence to export leadership. Failure to develop indigenous capabilities to produce specialized and globally competitive technologies is a major reason why many less developed economies get stuck in the middle-income trap \citep{lee2019art}. %\footnote{The middle-income trap refers to economies that rely heavily on natural resources and/or low-cost labor, and as a result, do not transition to a knowledge-based economy that produces high value-added products and services \citep{bresser2019classical}.}
Without developing such capabilities, countries often remain dependent on low value-added, labor- or resource-intensive sectors and cannot move to a knowledge-based economy with the ability to produce high value-added products and services \citep{fu2011role, bresser2019classical}.

Trends toward stronger and globally harmonized intellectual property rights (IPR) systems are among the factors that may have hindered the catch-up process \citep{chang2001intellectual, falvey2006intellectual, castaldi2024intellectual}. Intellectual property (IP) protection is intended to motivate innovation. 
However, obtaining IP protection through a patent requires that inventions meet high global standards of non-obviousness\footnote{Non-obviousness means that `the invention is not obvious to a person having ordinary skill in the field at the time of the invention.'} and novelty\footnote{Global novelty means that `the invention has not been publicly disclosed (i.e., is not part of the prior art).'}, meaning that an invention must be an advanced creation that shows a significant improvement over prior art. %, often by using laws of nature, \SJ{removed as it's same for the UM}
This excludes many incremental and adaptive innovations from patentability and imposes standards that are difficult to meet for inventors from less developed countries \citep{dreyfuss2021technological}. 
Moreover, the high level of technological sophistication required by patents often does not align with the domestic needs of developing countries, as most needs can be met by affordable and adaptive innovations based on existing technologies \citep{burrell2023intellectual}. These aspects make patents unsuitable for promoting innovation and active learning in the early stages of development \citep{viotti2002national}.

Other types of IPRs, such as trademarks and utility models (UMs), have been discussed as more conducive to promoting development and domestic entrepreneurship \citep{suthersanen2006utility, kim2012appropriate, gnangnon2014intellectual}. In particular, UMs have been suggested as an alternative to patents for inventors seeking domestic IP protection for incremental but locally useful technologies \citep{dreyfuss2021technological, cahoy2021legal}. UMs, or more generally, second-tier patents,\footnote{Many countries offer some kinds of second-tier IPR protection akin to UMs, but label them differently, for example, as `petty patents', `short-term patents', `utility registrations', `utility innovations', or `innovation patents' \citep{suthersanen2019utility}. We use the term UM when referring to second-tier patents more generally.% (see Section \ref{sec:background_UM}). %but acknowledge that empirical conclusions may be sensitive to specific design aspects of UM systems and their interaction with legal, socio-economic and technological contexts (see Section \ref{sec:Theory_hyp}).
} impose less stringent requirements for obtaining protection and often provide a cheaper and faster process than patents (see Section \ref{sec:background_UM} for details). This allows IP protection for imitative and incremental inventions that meet the domestic needs of less developed economies. \citet{kim2012appropriate} showed a positive relationship between firms' UM use and performance during the catch-up period, suggesting a beneficial role of UMs in the early stages of development. 

However, while there is discussion about the suitability of UMs in the development context \citep{kim2012appropriate, suthersanen2019utility, kang2020intellectual, burrell2023intellectual, heikkila2023key}, empirical evidence on their \emph{long-term} implications extending into the post-catch-up period is scarce. Moreover, despite their potential suitability, wider adoption and implementation of UM systems in developing countries has been limited. Given the diversity of catch-up strategies pursued by different countries, we need a better understanding of how and under what conditions UM systems can support technological learning during catch-up, and of their long-term implications. We address this gap by asking:  \textit{Can learning from imitative and adaptive innovations facilitated by a UM system during technological catch-up contribute to a country's ability to innovate competitively at the global frontier\footnote{See Section \ref{sub:3.1} for the definition of \emph{global frontier technology}.} in the post-catch-up phase? If so, what mechanisms enable the UM system to support learning in the catch-up process?} 
%Here we define \textit{global frontier technology} as a novel and non-obvious technology by global standards, with the potential to create value in international markets and export revenues \citep{sanders2018world}. %% KH: I commented the definition out as we introduce the definition later. I suggest to define it only one time

This study answers these questions by investigating how a catch-up strategy specializing in short-cycle technologies through imitative and adaptive innovation interacts with the design of the IPR system and UMs in particular. We focus on the case of South Korea (hereafter Korea), a country that has evolved from a catch-up economy to a global innovation leader in several high-tech industries in a few decades. We consider the United States Patent and Trademark Office (USPTO) patents of Korean applicants as indicators of the country's frontier technologies. Our results show that Korean frontier technologies tend to be more globally impactful (captured by forward citations) when they build on domestic UMs (captured by direct or indirect backward citations to UMs). This implies that UM-based active learning experiences, developed through incremental and adaptive innovations, can play a critical role in building globally competitive capability in the long run. Moreover, we show that the country's frontier technologies contribute more to local value-added relative to foreign value-added when they build on domestic UMs. This suggests the long-term relevance of UMs in internalizing value-added from frontier technologies, producing technologies that are distinct from those produced by other countries.

This study is one of the first to provide quantitative empirical evidence on the long-term impact of UM-based learning in the catch-up process. Our findings contribute in two main areas. First, we add to the literature on UM systems at different stages of economic development (e.g., \citealt{maskus1999impacts, kim2012appropriate}). We articulate and empirically underscore mechanisms by which a country's catch-up strategy may interact with the UM regime, leading to the development of core capabilities to produce specialized, high-impact, and local value-added frontier technologies in the post-catch-up period.  
Second, our findings are informative for IPR policy discussions, providing detailed empirical and theoretical insights into how the adoption of a UM system can function as a learning support mechanism in the catching-up process. Many developing countries do not have a UM system, and among those that do, the understanding of its effectiveness and long-term impact remains vague. 
Our results suggest that the traditional rationale for UM as a system to promote minor inventions tends to overlook the long-term implications, as UMs may influence the emerging industrial structure and competitive niche of a catching-up economy in the long run. We demonstrate the feasible benefits of UMs as part of the national innovation system (NIS) in catching-up economies, as well as the conditions under which such benefits are more likely to be realized.

The remainder of this study is structured as follows: Section \ref{sec:background_UM} introduces the UM system in Korea, and Section \ref{sec:theory:catch-up} derives our two hypotheses based on the technological catch-up literature and UM regime design. Section \ref{sec:Data and Analysis} describes the data and analyses. Section \ref{sec:Results} presents the results, Section \ref{sec:Discussion} provides context and discusses their implications, and Section \ref{sec:CM} concludes. 

\section{The Utility Model in Korea}
\label{sec:background_UM}
UMs exist in many countries but have received relatively little attention in innovation research compared to patents \citep{janis1999second, suthersanen2019utility, heikkila2023key}. Germany and the UK were the earliest to introduce UM systems in the mid-19th century, aimed at curing deficiencies in patent systems, especially in terms of cost, ease, and speed of application and protection. Compared to patents, UM applications are often cheaper and come with lower requirements regarding the inventive step, allowing minor and incremental inventions to be protected. This is intended to improve access to IP protection for inventors with limited technological and financial capacities. In many cases, simplified or lacking examination procedures further lead to significant reductions in the costs and time lag required to obtain an IPR for an invention.

Historical experiences with early UM systems are mixed: while the system in the UK was abolished early \citep{suthersanen2019utility}, the German model has been maintained \citep{koniger2017125th} and has become a role model that inspired similar systems around the globe. Prominent examples include Japan, China, and Korea, where UMs are often seen as a supportive mechanism for technological learning from imported technology through reverse engineering and adaptive innovation \citep{kim1997imitation, maskus1999impacts, kumar2003intellectual, suthersanen2006utility, huang2017institutional, kang2020intellectual}. Today, about 70 countries offer UMs or equivalent systems as a means of domestic IP protection \citep{heikkila2023key}.

The objectives of maintaining or introducing UM systems are similar across countries, yet the lack of international harmonization offers more flexibility to design UM systems in line with domestic needs, which has often been considered an important feature in the context of development \citep{heikkila2018need, dreyfuss2021technological, heikkila2023key}.\footnote{UMs are included in the Paris Convention, but beyond national treatment and priority, the Convention does not establish any standards to be met by UMs \citep{janis1999second}. Some jurisdictions allow an international UM application in line with the Patent Cooperation Treaty (PCT) \citep{wipo2023IPstatistics}.} This leads to a variation in UM systems across countries, often tailored to domestic needs and subject to changes over time in response to economic development, technological change, and international trade harmonization \citep{janis1999second, kim2015overview}.

Here, we focus on the key aspects of the Korean UM system and explain their differences in relation to patents. Many, but not all, of these comparative considerations are similar in other jurisdictions. Box 1 summarizes the three aspects that are most relevant for our study: (1) the ‘size’ of the inventive step, (2) the ‘scope’ of protectable subject matter, and (3) the ‘duration’ of protection. For the sake of completeness, we briefly discuss other features of the UM system that appear relatively less relevant in our context. 

\subsection{Size of the Inventive Step} 
Although both Korean patents and UMs are granted for inventions that are globally novel (i.e., not yet publicly known elsewhere), UMs require a less stringent inventive step than patents. UMs are granted for ``any device on the shape, structure, or combination of industrially usable articles'',\footnote{UM Act (1997): \url{https://www.law.go.kr/LSW//lsInfoP.do?lsiSeq=4614\&chrClsCd=010203\&urlMode=engLsInfoR\&viewCls=engLsInfoR\#0000}} while patents explicitly require an invention to be a ``highly advanced creation [...] utilizing rules of nature''.\footnote{Patent Act (1997): \url{https://www.law.go.kr/LSW//lsInfoP.do?lsiSeq=51815\&chrClsCd=010203\&urlMode=engLsInfoR\&viewCls=engLsInfoR\#0000}} 
The patent requirements exclude most incremental and adaptive inventions, which are often more realistic to achieve for inventors with limited financial and technological capacity. 
Beyond Korea, this aspect is often considered when justifying UMs as a means to improve the accessibility of IPRs to individuals, SMEs, and inventors from developing countries \citep{johnson2015economic, suthersanen2019utility, burrell2023intellectual}.\footnote{In other jurisdictions, the requirements for UMs are even lower, defining the standard of novelty to be met \emph{locally}, which enables the protection of imitative innovation and is thought to promote technology diffusion and local entrepreneurship \citep{dreyfuss2021technological}.} 
In a catch-up context, the lower inventive step requirements may be crucial, especially when active learning through incremental and adaptive innovation based on foreign technology is an integral part of a national development strategy \citep{viotti2002national}.

\subsection{Scope of the Protectable Matter} 
While UMs provide a tool for protecting incremental inventions, they are not eligible for the protection of methods, processes, and various chemical, bio, and medical inventions, unless these inventions relate to devices \citep{lee2014patents}. The limited scope of protection of the UM system focuses on product innovations in sectors with relatively short technology cycles, while excluding the majority of technologies characterized by long technology cycles, such as biotechnology, pharmaceuticals, and other advanced materials. Innovation in these long cycle sectors requires considerable experience in basic research. The patent system provides protection for these technologies. Material patents have been effective since the Korean patent law was amended in 1986, while process inventions were protected by patents prior to 1986 \citep{lee2010ipr, kim2015overview}.

\subsection{Duration of Protection} 
UM protection for an invention lasts for 10 to 15 years from the application date, which is shorter than the duration of patent protection, typically 20 years. In Korea, the duration of UM protection varied over time, from 12 to 15 years during the early catch-up period, and then became 10 years after 1999, when the country reached the maturing catch-up stage \citep{kim2015overview}. The short protection period of UMs is well suited for inventions with a relatively short technology cycle. In contrast, patents provide IP protection for inventors who seek to protect an invention over a long time horizon.

In other contexts, a faster time-to-grant is often described as another key difference between UMs and patents \citep{heikkila2018need}. While this is true in other jurisdictions where UMs are not subject to in-depth examination, the effect may be less significant in the Korean context, excluding the 1999-2006 period (see Section \ref{UM:1999-2006}). In other countries, the differences in the time lag from application to grant can be significant, with more than 1-2 years for patents, but only a few months for UMs. In Korea, both UMs and patents undergo a similar examination procedure, the length of which depends on the complexity of the application and the number of claims to be approved. Although there are no official statistics, anecdotal evidence suggests that the examination of UMs is faster than that of patents, which may be explained by the less rigid standard of patentability to be approved, especially when applications concern minor innovations with a limited number of claims.

\vspace{0.5cm}
\fbox{
    \parbox{0.93\textwidth}{
    \small
    \textbf{Box 1: Key features of Korean UMs compared to patents}
    \begin{itemize}
    \item \textbf{Less stringent patentability requirements:} UMs can be awarded for relatively minor improvements or adaptations based on existing technologies.
    \item \textbf{More rigid restrictions on protectable matter:} Process technologies, biotechnology, and materials are largely excluded, making the UM system more suitable for product innovations in relatively short-cycle sectors.
    \item \textbf{Shorter duration of protection:} 10-15 years in contrast to 20 years for patents. %; Both UMs and patents offer extension possibilities, but UMs cannot be renewed. 
    \end{itemize}
    References: \cite{lee2014patents}
    }
}
\vspace{0.5cm}

\subsection{Other Features of the UM System} Other features of UM systems that are often discussed in the context of development relate to the cost and complexity of filing, the lack of a globally defined novelty standard, and their accessibility to SMEs \citep{suthersanen2006utility, maskus2010differentiated, kim2012appropriate, heikkila2023key}. 

The difference in cost and complexity of filing a UM versus a patent is significant in many jurisdictions, especially those that do not require a detailed examination of a UM. This makes UM applications much cheaper, often costing only a few hundred USD, while the cost of filing a patent can be several thousand USD, including attorney fees. The examination requirements often increase the complexity of the application, which is an additional burden for inventors with limited technological and financial capacity \citep{suthersanen2019utility, heikkila2023key}. While UM filing costs in Korea are slightly lower than those of patents and tend to be associated with lower attorney fees, the differences are relatively small, as both UMs and patents follow the same examination procedures. However, there is a significant difference between domestic and international IP, as international IP can be very expensive to obtain. For example, the total cost of a USPTO patent application often exceeds USD 10,000-20,000, making these forms of IP inaccessible to inventors with limited capacity. 

Other studies also emphasized the possibility of defining the novelty standard locally instead of globally as a potential advantage of a UM system. This is thought to promote technology diffusion, imitative innovation, and domestic entrepreneurship \citep{suthersanen2006utility, dreyfuss2021technological}. This does not apply to the Korean UM system: it requires UMs to be globally new, which is validated by the examination.
%For novelty: maybe an issue in some development context and technology diffusion context (search the Japan paper) \citep{dreyfuss2021technological}. Not relevant here, because Korea has global novelty. Only in 1999-2006, they may have been a lot of imitative inno, but the system was abandoned exactly because of too many low quality cases. However, the novelty aspect may be relevant in other contexts, when diffusion not innovation is the objective (medicine, climate). 

Often, the easier accessibility of UMs, as opposed to patents, is seen as a means to stimulate innovation by SMEs. However, the empirical support for this objective is rather weak, as the main UM applicants remain large firms, although we often observe relatively higher participation of domestic and small firms compared to multinationals \citep{johnson2015economic, heikkila2018role, suthersanen2019utility}. 
This potential UM advantage is not a central consideration in our study, in part because SME support has not been a major consideration in Korean industrial policy, which has strategically promoted innovation-driven competition among large industrial conglomerates \citep{kim1997imitation}. This study focuses on the role of the UM system in helping these conglomerates catch up and move ahead of the global technological frontier in well-selected areas.

\subsection{The Korean UM System during 1999-2006}\label{UM:1999-2006} 
Since its inception in 1962, the characteristics of the Korean UM system and its relationship to patents have remained relatively stable over time, except for the period between 1999 and 2006, when the Korean government introduced a quick registration system for UMs.\footnote{Other reforms of the Korean UM and patent systems have been rather minor and involved adjustments, often in response to ongoing international harmonization efforts and technological changes. These include, for example, changes to the validity period, scope of protectable matter, and technical aspects of the application. For a comprehensive overview of the legal reforms in the Korean IPR system, see \citet{kim2015overview}.} 

During this period, in-depth examination was replaced by a simple formality check process, which reduced the time to grant from more than a year to a few months. The system also allowed dual applications, enabling a UM application to be converted into a patent application after examination. This enabled inventors to obtain strong and long-term protection through a patent while having their invention protected by a UM during the time they waited for the patent. The main objectives of this reform were to support technologies with a short cycle time and to support SMEs. As a side effect, the number of granted UMs surged during this period, with UMs granted for potentially low-quality inventions and cases of abuse. As a consequence, the system was abandoned in 2006, and the examination requirement was reinstated \citep{lee2014patents}. 

\section{From Catch-Up to Post-Catch-Up: A Learning Detour}
\label{sec:theory:catch-up}
Our study investigates how UM systems can help a catching-up economy build the necessary technological capabilities for the transition to the post-catch-up period. More specifically, we examine the long-term relevance of \emph{active} technological learning--supported by a UM system--during the catch-up period \citep{viotti2002national}, and its impact on innovation in the post-catch-up period \citep{abramovitz1986catching, soete1988catching, fagerberg2006innovation, lee2019art}.

Economic catch-up is defined as the closing of the productivity and income gap with developed countries \citep{fagerberg2006innovation}. Here, we focus on \emph{technological} catch-up, defined as closing the innovation capability gap between a latecomer and the technological leaders \citep{lee2017catch}, as a factor that enables a country to achieve high per capita income levels and a path of endogenous technology-driven growth \citep{romer1994origins}. Technological catch-up is often sector-specific, and its difficulty depends on the ease of learning and the lead time of incumbents. 
%Producing innovations in a diverse range of technologies requires high capability levels and a well-functioning supportive NIS. 

Countries that have developed innovation capabilities across a broad range of sophisticated, hard-to-imitate technologies are considered to have successfully achieved technological catch-up \citep{abramovitz1986catching, fagerberg2006innovation, hidalgo2007product, lee2021catching}. The process of how successful latecomers can achieve this capability level has been described as a \emph{detour} of technological learning, in which the order of prioritizing specific learning modes (Section \ref{sub:3.1}) and technologies (Section \ref{sub:3.2}) can be instrumental \citep{lee2019art, lee2021catching, lee2024economics}. 

This section explains how the UM system can be a supporting mechanism in the detour process to build the technological, financial, and institutional capacities required for technological catch-up and the transition to the post-catch-up phase. We develop two hypotheses about the supportive role of the UM system in enabling a country (1) to produce high-impact technologies at the global frontier and (2) to develop a unique specialization that allows it to generate localized knowledge spillovers from technologies that are hard for foreign inventors to imitate. 

\subsection{Imitation and Adaptation before Innovation}\label{sub:3.1}  Imitative and adaptive learning based on existing foreign technologies is a first step in the catch-up process, preceding the stage at which latecomers are able to create original technologies at the global frontier \citep{kim1997imitation, dobson2008transition}. 

Here, we define \textit{global frontier technology} as a technology that is novel and non-obvious by global standards and has the potential to create value in international markets and generate export revenues \citep{sanders2018world}. Patents registered in advanced economies, such as the United States, can serve as indicators of frontier technologies, as they require the patented invention to be globally novel and non-obvious, both of which are rigorously examined. Moreover, their filing and registration are expensive, indicating an expected commercial value of the invention in international markets.\footnote{Not every patent filed in an advanced economy leads to the development of commercial products. However, even when patents are not enforced or are filed for defensive reasons, they are indicative of the global technological frontier that reflects the edge of existing, codified technological knowledge.} 
Frontier technologies that are widely adopted tend to generate follow-on innovations and can be considered valuable, high-impact technologies \citep{trajtenberg1990penny, aristodemou2018citations}. % KH: I added the aristodemou reference, we can use it to justify our citation based measure as indicator of tech that creates much follow-on impact & commercial/economic value

The capability to produce frontier technologies is a symbol of advanced economies and is positively correlated with income levels and productivity. Catching up to the technological frontier is considered essential for economic development, as frontier technologies can be a major source of sustained value-added generation \citep{abramovitz1986catching, fagerberg2006innovation, baldwin2013trade}. They enable countries to produce and export (`non-ubiquitous') technologies and products that cannot be easily imitated \citep{hidalgo2007product}. 

However, achieving the capability level necessary to produce frontier technologies is not straightforward. Developing countries face many barriers, including a lack of tacit knowledge, limited production capacity, and weak institutional systems (e.g., finance, law, education, and standards) that support the transformation of codified knowledge into innovations \citep{sanders2018world}. Therefore, learning by doing and domestic experimentation, through imitating foreign technologies and producing locally useful adaptive innovations, can help countries gain experience in production and commercialization, develop supportive institutions, and establish a culture of innovation-driven business \citep{viotti2002national, acemoglu2006distance, chang2024dynamics}.

\subsection{Short-Cycle Technologies before Long-Cycle Technologies}\label{sub:3.2} However, success in imitative and adaptive learning is not guaranteed, and the general notion of `imitation' gives little guidance on the question of technology choice and specialization. In other words, which sectors and technologies should be selected first in an imitative learning process to maximize the probability of successfully catching up and, ideally, forging ahead? 

Previous studies have found that the probability of success is higher if a country prioritizes learning in sectors with a short technological cycle time (TCT)\footnote{TCT can be measured by the average (or median) age of the patents cited in a patent (or a group of patents) in a particular technology class \citep{kayal1999empirical}. The measure captures the immediacy of reliance on prior knowledge. Hence, in fields with long TCT, a technology tends to have long-lasting relevance for producing follow-on inventions \citep{lee2024economics}.}—such as electronic devices and telecommunications—before entering long TCT sectors, such as pharmaceuticals and biotech \citep{park2006linking, lee2019art, lee2021catching, lee2024economics}. A short TCT means a short lifespan of technologies in a particular field. By definition, a short TCT implies the rapid obsolescence of technological knowledge, which creates windows of opportunity for developing countries to enter knowledge production and commercialization \citep{soete1988catching, lee2017catch, chang2024dynamics}. Therefore, a short TCT provides a realistic chance of closing the gap, producing products at similar quality levels but lower costs than incumbents in high-income economies, and selling them internationally at competitive prices. Specialization and international commercialization in key short TCT sectors can deepen a country's learning experience in conducting domestic R\&D, trading in international markets, and building R\&D alliances with foreign partners.

%Further, short lead time can allow technological latecomers to eventually forge ahead \citep{lee2017catch}. 
%Therefore, the focus on short TCT increases the probability to succeed in the catch-up with international leaders.
%Finding a promising niche in the short-cycle areas is a critical matter of concern here.  
%However, since short TCTs come with rapid cycles of technological leadership change \citep{lee2017catch}, they may be risky as a basis when aspiring sustained technological leadership. 

In contrast, relatively high entry barriers exist in long TCT sectors, enabling leaders in these technologies to achieve high markups and generate value-added over a long period, with less pressure from competitive threats posed by new entrants \citep{petralia2017climbing, lee2023schumpeterian}. Innovation in these sectors tends to rely on extensive experience in applied and basic research, limiting the success of purely hands-on, imitation-based learning that requires relatively little pre-existing research capacity. 

Historical examples show that catch-up strategies through early investments in basic research and long-cycle sectors (for example, in Brazil and Mexico in the 1980s and 1990s) were unsuccessful due to a lack of capabilities to turn these investments into long-term value-added creation through innovation at the frontier, in competition with incumbents from advanced economies \citep{acemoglu2006distance, lee2024economics}. Transforming scientific advances into globally competitive products and services requires a supportive NIS, which is often underdeveloped in the early phases of technological catch-up \citep{nelson1985evolutionary, fagerberg2006innovation, lee2013schumpeterian}. Successful catch-up countries, such as Japan, South Korea, and Taiwan, prioritized learning focused on relatively short TCT sectors, such as semiconductors and electronic devices. Investments in basic research took off later, after these countries had established a certain level of technological, economic, and institutional capacity.

\subsection{Transition to the Post-Catch-Up Phase} 
The rapid cycles of technological obsolescence in short TCT sectors lead to a reduced reliance on knowledge stocks accumulated in advanced economies \citep{lee2021catching}. Rapid obsolescence requires new entrants to adapt their technologies to the new cycle themselves. Such \emph{indigenous} innovation capacity implies that entrants have developed a sophisticated and holistic understanding of why these technologies work and how to creatively modify them \citep{fu2011role}. Therefore, a detour via short-cycle technologies can help countries develop technological independence from foreign technology inputs and manage their entire innovation process. 

This independence forms the basis for developing a well-functioning NIS that supports subsequent technological \emph{diversification} \citep{lee2024economics}. An initial specialization in targeted short TCT sectors indicates not only technological learning but also the development of complementary assets, including physical infrastructure, human capital, and legal, financial, and cultural institutions. Successful initial specialization in short TCT sectors generates export revenues, which are essential for building financial capacity in the private sector for investments in follow-on innovations, and for establishing trade and R\&D relations at the international level. The incremental sophistication of the NIS brings positive spillovers for producing and innovating in related technologies \citep{nelson1993national, fagerberg2006innovation, hidalgo2007product}.  

Therefore, while \emph{specialization} in short-cycle sectors continues, the country can gradually \emph{diversify} its industrial structure in short TCT sectors \citep{lee2021catching}. Diversification is reflected in the diversity of technology fields in which the country innovates. In addition, specialization and diversification within short TCT sectors are associated with a declining average TCT of a country. The average TCT of USPTO patents filed by several successful latecomer countries has declined during the phase of rapid catch-up, followed by a few years of stagnation and a subsequent slow rise \citep[see Fig. 1 in][]{lee2021catching}, while still remaining at lower levels than those of advanced economies and countries such as Brazil, Mexico, or South Africa, which prioritized long-cycle technologies in their industrial policies \citep{lee2024economics}. 

%Innovation in highly sophisticated technology areas is a supporting condition for the transition to high-income status. \citep{hartmann2021did}. 
Countries with diversified capabilities have more options to combine capabilities across different areas. This is a key source of sophisticated and non-ubiquitous innovation at the frontier \citep{hidalgo2007product, lee2021catching}, as well as of the transition to high-income status \citep{hartmann2021did}. Accordingly, the process of specialization and diversification in short-cycle technologies enables a catching-up economy to build its indigenous capabilities and establish its core industrial structure, which holds the potential to be competitive and unique in the global market during the post-catch-up period.

%/*revise - note that not all countries' UM design is like this (R1's comment). Link our discussion better to the Korea's UM context that facilitated imitative and adaptive learning*/
%\url{http://www.koreanlii.or.kr/w/index.php/Utility_model_right#:~:text=A%20utility%20model%20right%20(%EC%8B%A4%EC%9A%A9,and%20less%20stringent%20patentability%20requirements.} and \url{https://www.kipo.go.kr/en/HtmlApp?c=20302} and \url{https://www.kipo.go.kr/en/HtmlApp?c=92004&catmenu=ek03_04_01} and \url{http://www.koreanlii.or.kr/w/index.php/Patent_cases}.
\subsection{The Supporting Role of UMs during the Catch-Up Phase} 
A well-functioning IPR system is an important component of an efficient NIS that supports the production, diffusion, and commercialization of economically useful knowledge \citep{nelson1993national}. Different types of IPRs exist, and their role in supporting innovation varies between technologies and levels of technological capability within a country \citep{kim2012appropriate}. Early access to appropriate forms of IPR protection is relevant for an innovation-driven catch-up, as it provides economic incentives to invest in technological learning, generates financial resources for private investments in follow-on innovation, increases IPR awareness, and helps establish a culture of innovation-driven business \citep{kim2012appropriate, dreyfuss2021technological}. UM systems are particularly well suited to support an innovation-driven catch-up strategy that follows the learning detour described above. 

The first steps in the detour are characterized by \emph{active} learning through imitation and technological experimentation, with adaptive innovation based on foreign technology \citep{viotti2002national}. Patents are less useful in this process because high standards of patentability prevent adaptive inventions and minor improvements to existing technologies from being protected \citep{dreyfuss2021technological, khouilla2024does}. UMs offer an alternative here, as they impose lower requirements on the size of the inventive step. In Korea, a UM can be granted for a relatively marginal and adaptive innovation as long as it is novel (i.e., not yet publicly known elsewhere) and represents a useful improvement over an existing, publicly known solution.

The second element of the detour relates to the prioritization of short- over long-cycle technologies in the catch-up process. As short-cycle technologies have a short lifetime, there is little added value in having long-term IPR protection provided by patents. UMs provide an alternative here, as the protection period is relatively shorter than that of patents. In addition, the rapid pace of innovation in short TCT sectors creates intense competition, which can create \emph{a need for speed} in obtaining protection \citep{heikkila2018need}. UM application and registration processes tend to be less complex and less expensive than those of patents due to lower inventive-step requirements, which reduces the time gap and resources to obtain protection. Moreover, the scope of protectable matters of UMs is well suited to protect inventions in short TCT sectors, which are represented by electronic devices and other engineering products. These features of UMs provide appropriate conditions to protect against threats from domestic competition in short TCT sectors, offering economic incentives and helping generate financial capacity for private R\&D. 

Overall, the key characteristics of the UM system are complementary to catch-up strategies that take a learning detour with an initial specialization in short TCT sectors. During the transition to the post-catch-up phase, specialization and diversification continue at the global frontier, now encompassing not only short-cycle but also long-cycle technologies, such as biotechnology and pharmaceuticals \citep{lee2021catching}. Despite ongoing progress in diversification, the industrial structure and capabilities developed throughout the catch-up phase are likely to remain influential in post-catch-up innovation, since technological evolution is path dependent \citep{nelson1985evolutionary}. The established industrial structure and capabilities are likely to form the foundation and knowledge base for follow-on innovation at the frontier. Therefore, we arrive at the following hypothesis: 
\begin{hypothesis}
    Frontier technologies produced by a post-catch-up country are more impactful when they build on the country's prior imitative and adaptive innovations treated as UMs during catch-up.
\end{hypothesis}

One of the important characteristics of a successful catch-up economy is a high level of knowledge \textit{localization} \citep{lee2019art, lee2021catching}, which is relevant in two ways. First, knowledge localization refers to the extent to which a country's innovation relies on knowledge spillovers from local rather than foreign sources. High localization implies relatively higher knowledge spillovers from domestic innovations to domestic, rather than to foreign, follow-on innovations \citep{jaffe1993geographic}. Since knowledge spillovers reflect the public-good nature of innovation and are a key driver of endogenous economic growth \citep{romer1994origins}, high levels of \emph{localized} spillovers indicate a country’s ability to generate endogenous innovation-driven economic growth from its indigenous capabilities \citep{jaffe1993geographic, romer1994origins}. 

Second, knowledge localization implies constraints on the extent to which spillovers are global, reflecting the need for technology-specific absorptive capacities to make creative use of knowledge. In that sense, knowledge localization signals a country's technological \emph{specialization} and, when associated with global frontier technologies, its competitive advantage in high-value-added production and trade. Therefore, localization reflects a country's ability to produce differentiated and hard-to-imitate innovations, enabling it to export unique high-value-added products in the global market \citep{jaffe1993geographic, hidalgo2007product, lee2018global}. 

Since innovation in the early stages of development is characterized by a strong reliance on foreign knowledge, an important question for the post-catch-up transition is how a country can achieve a high level of knowledge localization in the production of frontier technologies, internalizing value-added and generating indigenous knowledge spillovers for sustained growth. 
A learning focus on short TCT sectors during catch-up can form the basis for knowledge localization. Although a country may begin by imitating foreign technologies, successful catch-up involves developing unique experiences and capabilities that differentiate it from foreign competitors. During the process of specialization and diversification in short TCT sectors, learning goes beyond reverse engineering and imitation; it includes mastering the skills needed to adapt to local markets and develop useful new products, while gradually relying less on foreign knowledge \citep{lee2021catching, lee2024economics}. 

As the focus on short TCT sectors requires new entrants to \emph{rapidly} accumulate such experience, this learning process can be supported by the UM regime. The UM system is designed to encourage incremental but useful variations of devices and to provide a speedy legal process. Moreover, the novelty requirement in the Korean UM system demands that inventors remain creative and seek a deeper understanding of technologies, as pure reverse engineering is not protected.\footnote{As discussed in Section \ref{sec:background_UM}, other jurisdictions often apply less stringent requirements by defining novelty at the national rather than the international level, which allows purely imitative inventions to be protected under UM systems.} These characteristics are well suited to facilitate indigenous innovation and technological independence, which are critical in short TCT sectors for timely self-adaptation to new cycles. It is plausible to expect that these key features of the UM system can support the knowledge localization process, subsequently contributing to internalizing value-added from the country's knowledge production. Therefore, we speculate that the core structures and capabilities shaped through UM-based learning in short-cycle technologies become a principal source of post-catch-up knowledge production that promotes local value-added relative to foreign value-added through localized knowledge spillovers.
 
%Prioritized learning in short TCT sectors with the UM system can help developing countries facilitate knowledge localization, accumulating tacit knowledge for innovative and unique adaptation, which is necessary to survive in a rapidly changing market. 
%- knowledge localization (national self-citation) reflects upgrading (p. 97).

\begin{hypothesis}
Frontier technologies produced by a post-catch-up country contribute more to local value-added relative to foreign value-added when they build on the country’s prior imitative and adaptive innovations treated as UMs during catch-up.
\end{hypothesis}

\begin{figure*}[htp]
%\centering
\includegraphics[width=11cm, height=12cm,keepaspectratio]{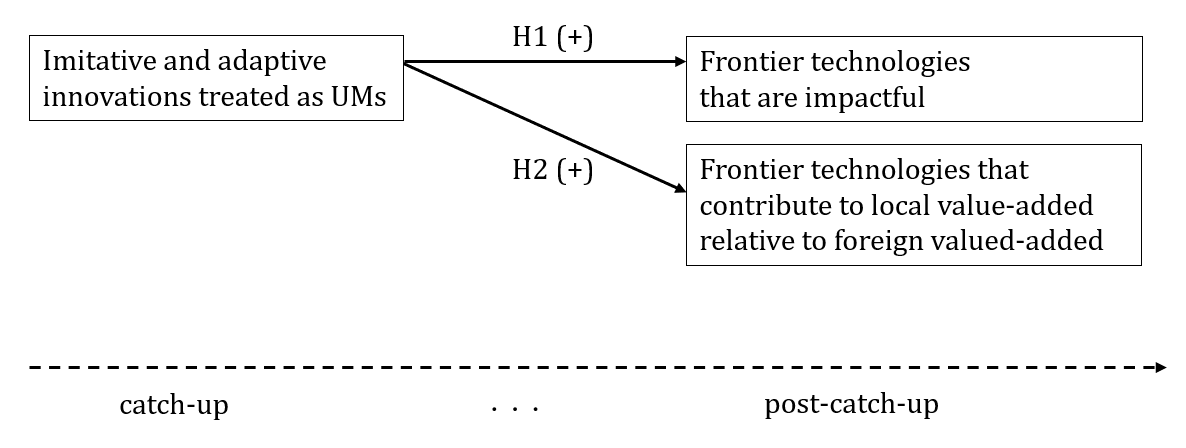}
\caption{Conceptual framework}
\begin{flushleft}
\begin{minipage}{1\linewidth}
\footnotesize
%\emph{Notes. }{*UM reliance: Direct or indirect connection to UMs in the citation network} 
\end{minipage}
\end{flushleft}
\label{FIG0}
\end{figure*}

\section{Data and Analysis}\label{sec:Data and Analysis}
\subsection{Case Introduction}\label{subsec:case}
Korea is one of the few countries that escaped the middle-income trap, transitioning from a developing country largely based on agriculture in the 1960s to a high-income nation within a few decades \citep{gill2007east, lee2019art}. %Korean firms nowadays successfully compete with global leaders in various markets, ranging from consumer electronics and machinery to integrated circuits, motor vehicles, ships, nuclear energy, and polymers, whose production relies on complex systems and associated know-how \citep{lee2015comparative, kwak2020unpacking, park2020evidence}. 
Korea's successful transition from \emph{imitation} to \emph{innovation} benefited from industrial policy and public investment in human capital and strategic industries, coupled with political pressure to maintain intense competition among the major firms known as \emph{chaebols} \citep{kim1997imitation}. %The government nurtured a system of oligopolistic competition between chaebols, supporting them in a carrot-and-stick manner. The support consisted of favorable regulations, financial benefits, and strategic investments in complementary public research and education, but multiple corporations competing in the same market benefited. This created an environment of intense oligopolistic competition, with strong R\&D incentives. For example, the policy pushed large firms to produce domestic prototypes (e.g., microwaves, cars, etc.) within a short time, while always being threatened by domestic competitors in a similar industry \citep{kim1997imitation}. 

From the mid-1970s to the mid-1980s, the country was in the \emph{beginning of the catch-up} period, during which foreign technologies were actively imported and imitated \citep{lee2010ipr}. During this period, the role of public support and industrial policy was crucial. The government nurtured a system of \emph{active learning} by supporting the chaebols in a carrot-and-stick manner. They benefited from favorable regulations, financial support, and strategic investments in complementary public research and education. Yet, these benefits were provided to multiple firms competing in the same market, creating intense oligopolistic competition. This environment encouraged active learning and the accumulation of hands-on-experience through R\&D races based on imported technologies, reverse engineering, prototyping, scale-up, and market experimentation \citep{kim1997imitation, viotti2002national, chung2015absorptive}.

These conditions constituted a supportive NIS, which enabled the country to enter a \emph{rapid catch-up} period from the mid-1980s to the mid-1990s. During this period, the private sector began to establish in-house R\&D centers and engage in science-based R\&D activities \citep{lee2010ipr}. Large firms increased their production activities in knowledge-intensive sectors, especially in relatively short TCT fields, including electronic devices, semiconductors, and mechanical engineering \citep{kim1997imitation, chung2015absorptive}. The UM system was part of the NIS, aimed at supporting a market-driven technological learning process (see Section \ref{sec:background_UM}) \citep{lee2014patents, kang2020intellectual}. The use of the UM system increased steadily during the early and rapid catch-up period (see Figure \ref{Figure3b}). 

The mid-1990s to the mid-2000s can be interpreted as a \emph{maturing catch-up} period, when Korean firms started patenting at the USPTO. This signals the beginning of frontier technology production in the global market (see Figure \ref{Figure3a}). During this period, Korea also increased its investment in long TCT sectors, such as biotechnology, pharmaceuticals, and advanced materials, to diversify its technological capabilities \citep{lee2010ipr}.            

\subsection{Data}\label{subsec:data}
We use the PATSTAT Global database (2023 Spring version), which covers USPTO patents and their citations, and UM data from the Korean Intellectual Property Office (KIPO) since the mid-1970s. 
The USPTO citation information is complete and includes citations to prior art, including Korean domestic UMs. While some details of UMs (e.g., abstracts) are incomplete, basic information such as application ID is fully available for the relevant period.\footnote{\url{https://public.tableau.com/app/profile/patstat.support/viz/CoverageofPATSTAT2023SpringEdition/CoveragePATSTATGlobal} [last accessed 01/04/2024]}%\footnote{\KH{Clarification: what are "application codes"? appln id or CPC codes or anything else?}} 

To capture frontier technologies produced by Korean entities, we collect USPTO-granted patents (hereafter, US patents) filed by Korean entities during 1976-2022.\footnote{The database covers patents published until February 2023.} Patents granted by the USPTO are often used as a proxy for the technology frontier because the US is the largest market in the world for many technology-intensive sectors, and its IP regime is among the strongest globally \citep{granstrand2018evolving}. We consider only granted patents and UMs to focus on valid inventions and analyze patents at the DOCDB family level to avoid double counting the same inventions. Table \ref{TableA1} highlights the 10 major players leading Korea's frontier technology production, indicating a high concentration among a handful of large conglomerates, which are commonly considered the drivers of Korea's technological catch-up \citep{kim1997imitation}. 

%We further collect forward and backward citation information of these patents (i.e., US patents applied by Korean entities).\footnote{\KH{To clarify before I do any rewrite: We further collect backward and forward citation information of those USPTO patents that are filed by Korean entities. $\rightarrow$ TBC: Hence, we do not collect citation info in general, right? But we collect citations \& the meta information of citing patents (esp. the year of citation event (filing or grant of citing patent?) and the assignee country), right?}} 
%Backward citations are used to compute a patent's reliance on Korean domestic UMs (see Section \ref{Ind}) and forward citations reflect the impact of a patent on follow-on innovations \citep{hall2000market, jaffe2000knowledge, jaffe2019patent}. We use forward citations to construct the target variable in our model to test hypotheses (see Section \ref{dv and model}).\footnote{\KH{Can we be more precise? Target variable = cumulative citation count or CDF (cumulative distribution function)? Note: for RP, we should not be too technical with the wording but precise.}}

\subsection{Dependent Variable and Model}\label{dv and model}
Our hypotheses concern whether, and to what extent, UM system-based learning helps a catch-up economy build its capacity to produce high-impact frontier technologies in the post-catch-up period (H1) and to internalize the value-added generated from these technologies (H2). %local value-added (relative to foreign value-added) (H2) frontier technologies in the post-catch-up period.\footnote{\KH{Proposed alternative wording for H2: ... to produce high-impact frontier technologies in the catch-up period (H1), and to internalize the value added generated from these technologies (H2). --- Idea behind the wording: value-added is difficult wording wise; with "internalization" of value added, we capture teh "relative to foreign" aspect and we also capture the "internalization of positive knowledge externalities" dimension. See endogenous growth theory (cited above) \citep{romer1994origins}. Let's discuss the wording again; I am not strong in my suggestion.}} 
We use granted US patents filed by Korean entities as a proxy for the country's frontier technologies and use forward citations to measure their impact on follow-on inventions \citep[e.g.,][]{trajtenberg1990penny, jaffe2000knowledge, hall2000market, jaffe2019patent}. We consider only the citations received from granted US patents, as they likely indicate more novel and valuable follow-on inventions than non-granted ones (see Section \ref{sub:3.1}). 

\begin{figure*}[htp]
\centering
\includegraphics[width=13cm, height=13cm,keepaspectratio]{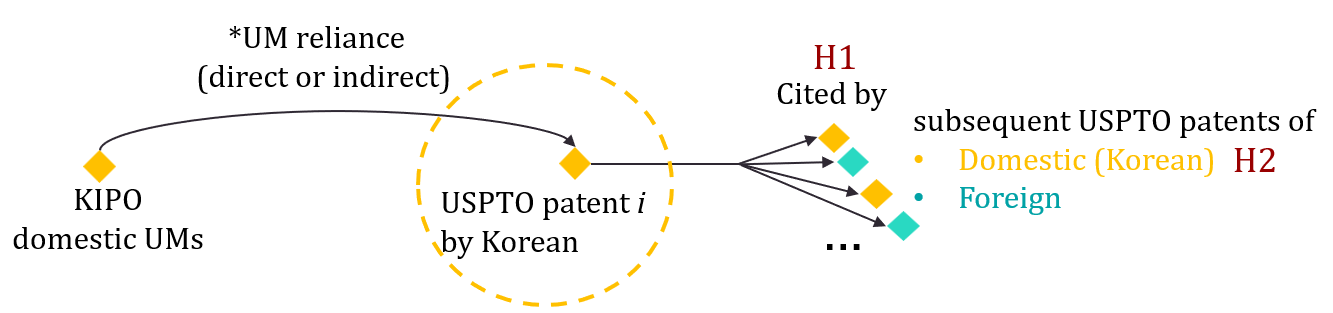}
\caption{Research design}
\label{FIG0}
\begin{flushleft}
\begin{minipage}{1\linewidth}
\footnotesize
\emph{Notes. }{*UM reliance: Direct or indirect connection to UMs in the citation network} 
\end{minipage}
\end{flushleft}

\end{figure*}

Following previous studies that model repeated forward citation events \citep[e.g.,][]{podolny1995role, nerkar2005evolution, jee2023firm}, we use a recurrent event hazard rate analysis that extends Cox regression to repeated events and allows for time-varying covariates. Compared to count models that use the frequency of \textit{n}-year forward citations as the dependent variable, the recurrent event model offers two major advantages that are relevant in our context \citep{kalbfleisch2002statistical, cook2007statistical}.\footnote{For H1, we conduct additional robustness checks using count models with 5-year forward citation frequencies as the dependent variable (see Section \ref{subsec:robustness_checks} and Table \ref{rob_count}).} First, it allows us to distinguish between different types of citation events within a single model. In our case, we can distinguish follow-on inventions made by Korean domestic actors from those made by foreign actors, allowing us to test our H2. Second, while the count model considers forward citation events within \textit{n} years, the recurrent event model captures long-term forward citation patterns until the end of the observation period. This feature provides flexibility in capturing long-term citation dynamics and reflecting time-varying conditions. The regression equation takes the following form:

$$ \lambda_i(t) = \lambda_0(t) \exp(\beta z_i + \gamma x_i(t)) $$

where $\lambda_i(t)$\ is the forward citation rate of a Korean US patent $i$ from time $t$ to $t + \Delta t$; $\lambda_0(t)$\ is a baseline rate that does not have any assumption about its distribution;
$x_{\text{i}}(t)$\ and $z_{\text{i}}$\ indicate vectors of time-varying and time-invariant covariates, respectively. The dependent variable for modeling the forward citation rate is given by the time interval between forward citation events of a patent $i$: the first forward citation is used as the first event, and then the intervals between subsequent citation events are sequentially used to model the repeated events. Robust standard errors clustered by each patent $i$ is used to account for multiple citation events per patent.%\footnote{\KH{To be sure that I understand this: errors are clustered because there is heteroskedasticity across patents dependent on the absolute number of citations received by a patent?}} 

\subsection{Independent Variables and Hypotheses}\label{Ind}
Our key explanatory variable aims to capture whether a patented invention builds on learning experiences that were facilitated by the domestic UM system. We construct a variable indicating whether a Korean US patent $i$ relies on the prior knowledge encoded in a UM granted by KIPO. This measure is obtained from a citation network based on direct and indirect citations from Korean US patents to Korean domestic UMs (see Appendix B for details on network construction). Based on the network, we compile a binary variable, \textit{UM reliance}, which takes the value of 1 if patent $i$ is directly or indirectly related to an earlier Korean domestic UM, and 0 otherwise. Based on this, the baseline regression model for testing H1 can be written as follows:

\begin{eqnarray}
    \textbf{H1: } \lambda(t) = \lambda_0(t) \exp(\beta_1 \textit{UM reliance} + \beta_{ctrl} \textit{Controls} + \gamma_{ctrl}\textit{Controls}_{t}) \,
    \label{eq:H1}
    \vspace{2mm}
\end{eqnarray}

To assess the extent to which \emph{UM reliance} contributes to more impactful (more frequently cited) patents filed by Korean entities (H1), we evaluate the coefficient $\beta_1$. H1 would be supported if $\beta_1$ enters with a positive coefficient. %As a robustness check, we also construct a distance measure that computes the minimum distance between the US patent and Korean UM in the citation network, counting a direct citation link as a unit distance (Appendix B and Table \ref{rob_count}). For example, if a patent cites a patent that cites a UM, it would be weighted with a distance of two, counting the steps from the focal patent via another patent to the UM. 

%Next, we examine whether the contribution of the UM reliance to a higher impact decreases with the country's level of economic development (\emph{Hypothesis 2}). We use the year of the first filing of patent $i$ at the USPTO ($Year$) as a proxy for the level of development, given the strong positive correlation between time and the development of the Korean economy since the 1970s (see Appendix A).\footnote{We checked that replacing $Year$ to the yearly $GDP$ gives the same direction of results. We cannot include $GDP$ and $Year$ in the same model due to multicollinearity, arising from high correlations between the two variables.} The regression specification to test H2 is:

%\begin{eqnarray}
%    \begin{aligned}
%    \textbf{H2: } \lambda(t) = \lambda_0(t) \exp(\beta_1 \textit{UM reliance} + \beta_2 \textit{Year} + \beta_3 \textit{UM reliance×Year} \\+ \beta_{ctrl} \textit{Controls} + \gamma_{ctrl} \textit{Controls}_{t}) \,
%    \end{aligned}
%    \label{eq:H2}
%    \vspace{2mm}
%\end{eqnarray}

%H2 would be supported if ${\beta_3}$ enters with a negative sign, indicating that time moderates the impact of the UM. %and leading to lower citation rates of UM-reliant patents compared to non-UM-reliant patents. 

In H2, we argue that UM-based learning experiences lay the foundation for technological specialization and future frontier innovation activities that contribute to value internalization via localized knowledge spillovers. To examine this, we use a dummy variable, \textit{Home citation}, that equals 1 if a forward citation was made by a Korean entity and zero if it comes from a non-Korean (i.e., foreign-based) entity. In our data, we observe 668,650 home and 1,748,199 foreign citation events. The model for testing H2 is given by:

\begin{eqnarray}
    \begin{aligned}
    \textbf{H2: } \lambda(t) = \lambda_0(t) \exp(\beta_1 \textit{UM reliance} + \gamma_1 \textit{Home citation}_{t} + 
    \gamma_2 \textit{UM reliance×Home citation}_{t} \\+ \beta_{ctrl} \textit{Controls} + \gamma_{ctrl}\textit{Controls}_{t}) \,
    \end{aligned}
    \label{eq:H2}
    \vspace{2mm}
\end{eqnarray}

H2 would be supported if ${\gamma_2}$ enters with a positive sign, indicating that UM-reliant patents tend to receive a higher share of citations from Korean domestic entities relative to foreign entities, compared to non-UM-reliant patents. A higher share of citations from domestic entities is indicative of Korea's technological specialization and competitive advantage: knowledge encoded in patents has public good characteristics with positive externalities that can only be absorbed and leveraged for future innovation if a country has developed technology-specific absorptive capacities. We interpret this as an indicator of local value-added relative to foreign value-added generation, whereby citations reflect the direction of follow-on technological development. Our models consider the censored time for each patent to be the last day of 2022, resulting in a total of 2,959,879 records (censored records are considered for home and foreign cases, respectively), with 2,416,849 forward citation events for our main regression (Table \ref{main}).

\subsection{Controls}
We control for additional factors that may affect the technological impact of a patent $i$. While the UM and patent systems differ in many respects (see Section \ref{sec:background_UM}), they share features in that they both aim to incentivize innovation through legal IPR protection. The two systems complement each other but can also compete in a domestic context, especially during the period when both are actively used. To reflect this, we control for the reliance of Korean USPTO patents on domestic Korean patents. Analogous to our \emph{UM reliance} indicator, we include a dummy variable indicating whether a Korean USPTO patent is linked (directly or indirectly) to KIPO patents (1) or not (0) (\emph{Domestic patent reliance}). %\footnote{Another concern may be the dual filing of domestic and/or international patents and UMs for the same invention. Since dual filings of Korean UMs and international patents were rare during the early development phase (see the fourth paragraph of Section \ref{subsec:Results-descriptives} for relevant statistics), it can be assumed that dual filings played a minor role in our context. The UM system probably played a role that could not be easily substituted by domestic patents.} 
This variable is constructed using a similar approach to that used to measure \textit{UM reliance} (see Appendix B), but based on a network of citations from Korean US patents to Korean domestic patents (instead of UMs).

Korea's economic development has largely been driven by a few giant conglomerates, such as Samsung, LG, Hyundai, and SK, which have been strategically supported by the government \citep{kim1997imitation, lee2001technological}. 
The majority of companies that have managed to accumulate technological capabilities to innovate at the global technological frontier were among these companies. Table \ref{TableA1} shows that 9 out of the top 10 Korean entities (excluding the Electronics and Telecommunications Research Institute (ETRI)) patenting in the USPTO are the chaebols. The contribution of the top 10 entities accounts for 88\% of the USPTO patents used in our analysis. The remaining 12\% come from various other applicants, including SMEs, universities, public research institutions, and individuals. One might expect that the level of technological accumulation in large applicants might have a more significant impact than that in other distributed entities. Therefore, we include a dummy variable indicating whether a patent's applicants include the top 10 largest entities (1) or not (0) (\textit{Major applicants}).

The diffusion of a new idea, product, or service often follows an S-curve. Therefore, the forward citation rate of a patent $i$ could be affected by its age. Previous studies using recurrent event models have controlled for the age effect \citep{podolny1995role, nerkar2005evolution}. Following these approaches, we consider the age of patent $i$ in time $t_{\text{i}}$ when a citation event occurs. This is measured as the time gap between the forward citation event and the filing date of the focal patent $i$ (\textit{Cited citing gap}) and its squared term (\textit{Cited citing gap squared}). In addition, to control for the time-constant effects of unobserved factors that may affect the propensity of a patent to be cited \citep{heckman1980does, nerkar2005evolution, marco2007dynamics}, we use the number of forward citations to patent $i$ that occurred on patent $i$ before a new citation event occurs in $t_{\text{i}}$ (\textit{Num prior cit}).

We also control for focal patent $i$-specific characteristics known to be related to patent impact. These include the scope of the patent, proxied by the number of claims (\textit{Num\_claims}) and references (\textit{Num\_references}), the size of the patent family (\textit{Family\_size}), and team size (\textit{Num\_inventors}) \citep{lerner1994importance, neuhausler2011patents, breitzman2015inventor} (see Table \ref{cor} for the correlation matrix). In addition, we include technology field dummies using the primary CPC section information of each patent $i$ (\textit{Field effects}) to account for different citation patterns across sectors. Finally, we include year effects to control for differences in the years in which patent $i$ is created and to better reflect the truncated nature of forward citation event data (\textit{Year effects}).
\FloatBarrier
\section{Results}
\label{sec:Results}
\subsection{Descriptives}
\label{subsec:Results-descriptives}
\begin{figure*}[htp] \centering
\begin{subfigure}{0.615\textwidth}
\includegraphics[width=10.2cm, height=9.2cm]{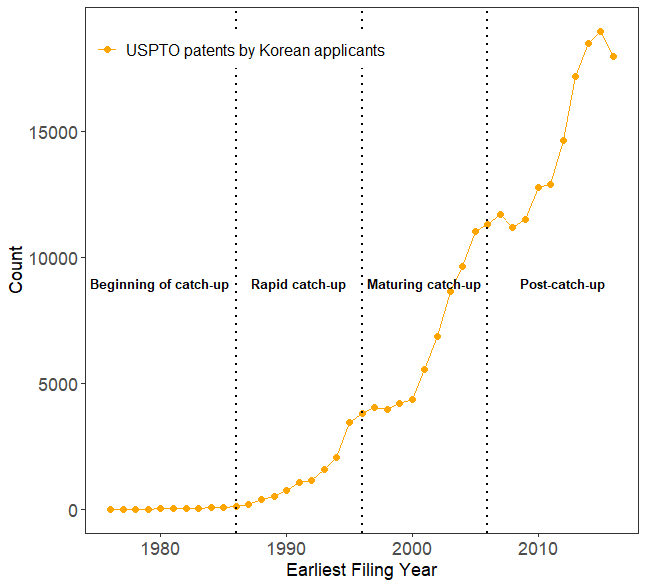}
\caption{\centering USPTO patents by Korean applicants}\label{Figure3a}
\end{subfigure}
\vskip 1em
\begin{subfigure}{0.8\textwidth}
\centering
\includegraphics[width=10.2cm, height=9.2cm]{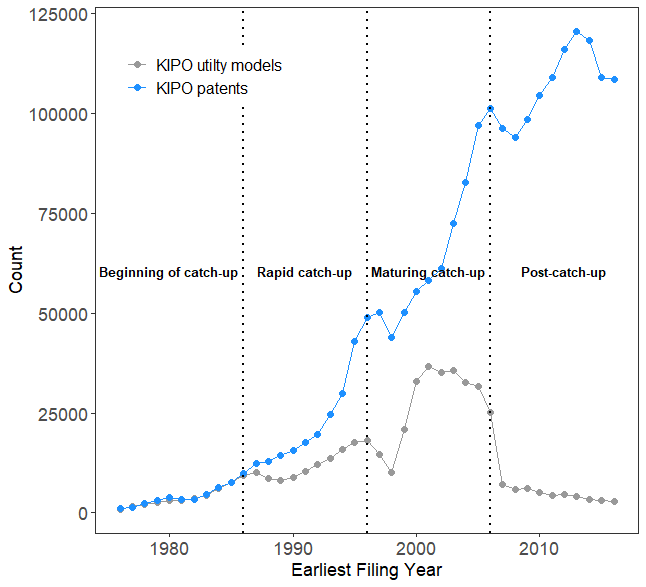}
\captionsetup{justification=centerlast}
\caption{\centering KIPO utility models and patents}\label{Figure3b}
\end{subfigure}
\vspace*{1mm}
\captionsetup[subfigure]{font=Large,labelfont=scriptsize}
\caption {Overall trends of inventive activities}\label{FIG:overall_inventive_activities}
\begin{flushleft}
\begin{minipage}{1\linewidth}
\footnotesize
\emph{Notes. }{Figure \ref{Figure3a} and \ref{Figure3b} show the annual number of granted USPTO and KIPO patents and KIPO UMs of Korean applicants by the earliest filing year, counted at a family level.
%Figure \ref{Figure3a} and \ref{Figure3b} present granted IPRs by the earliest filing year, counted at a family level.
} %Yellow, gray, and blue lines correspond to USPTO patents by Korean entities, KIPO UMs, and KIPO patents, respectively.} %Figure \ref{Figure1b_UM_reliance} illustrates that the proportion of frontier knowledge production relying on UMs has consistently increased, despite the declining number of UMs in Korea since around 2006 (gray graph in Figure \ref{Figure1a_overall}). Direct reliance has remained relatively consistent, while indirect reliance has steadily increased over time.} 
\end{minipage}
\end{flushleft}
\end{figure*}

Figure \ref{FIG:overall_inventive_activities} shows the long-term trends of inventive activities related to our analysis. The figures distinguish different phases of the Korean catch-up process, relying on \citet{lee2010ipr} and the descriptions introduced above (Section \ref{subsec:case}). 

Figure \ref{Figure3a} shows a sharp increase in the number of Korean USPTO patents since the 2000s, indicating a growing involvement of Korean entities in frontier technology production. %A few major companies have driven the country's production of frontier knowledge (see Appendix Table \ref{TableA1}). 
The gray graph in Figure \ref{Figure3b} shows increasing use of UMs granted by KIPO until the mid-1990s, followed by a slight decline and a sharp increase after 1999, then a steep decline and stagnation after the mid-2000s. The decline and increase between 1999 and 2006 may be related to a major reform of the Korean UM system. In 1999, KIPO suspended the examination process and introduced a quick registration system to promote the development and commercialization of short-cycle technologies. The quick registration system was abandoned for UM applications after 2006 (see Section \ref{UM:1999-2006}).\footnote{See \url{https://www.kipo.go.kr/en/HtmlApp?c=92002&catmenu=ek03_01_02_01}.} In contrast, we observe a growing number of KIPO-granted patents until recently (blue graph in Figure \ref{Figure3b}).

\begin{figure*}[htp]
\centering
\includegraphics[width=10cm, height=10cm,keepaspectratio]{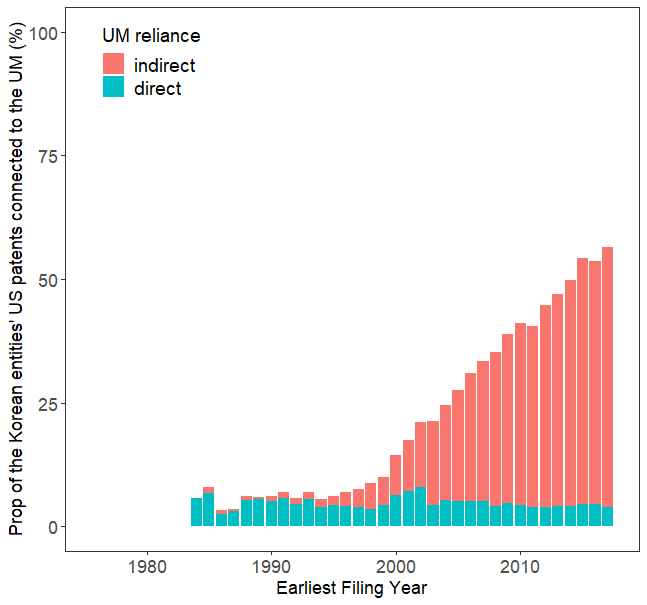}
\caption{Korean US patents relying on UM (\%)}
\begin{flushleft}
\begin{minipage}{1\linewidth}
\footnotesize 
\emph{Notes. }{This figure shows that the proportion of frontier technology production relying on UMs has consistently increased, despite the declining number of UMs in Korea since around 2006 (gray graph in Figure \ref{Figure3b}). Direct reliance has remained relatively consistent, while indirect reliance has steadily increased over time.} 
%The `high value' (gray graph) here denotes the patents cited at least once by other US patents %\KH{filed by Korean entities?} %within 5 years of application. Corresponding trends for Korean domestic patents are illustrated in Figure \ref{FIG_A1}, revealing a peak in the graphs approximately ten years later compared to that of UMs.%\KH{Does this figure include indirect citations?}  
\end{minipage}
\label{FIG4}
\end{flushleft}
\end{figure*}

Figure \ref{FIG4} shows the proportion of Korean US patents based on UMs. This proportion has been steadily increasing despite the rise and fall in the number of UMs in the early 2000s (Figure \ref{Figure3b}). The proportion of patents directly relying on UMs has remained almost constant for decades (around 5\%),\footnote{The stability of patents directly relying on UMs, regardless of the unstable evolution of annual UMs, suggests the possibility that the reform of the Korean UM system in 1999 led to a higher number of low-quality UMs that would not have passed proper examination, while the number of UMs that contributed to a knowledge base for subsequent global frontier innovation was almost unaffected by the reform.} while those with indirect UM reliance have increased steadily (reaching $>$50\% in the late 2010s). This suggests increasing technological specialization: more and more patent-protected inventions build on the country's earlier inventions, which in turn build on inventions originally protected by UMs. UM-protected technologies can be seen as the starting point in a chain of learning through follow-on inventions that are eligible for US patent protection.

The Korean UM system allows dual filings, i.e., the same inventor may file a domestic UM and a domestic and/or international patent simultaneously. Dual filings may be indicative of strategic uses of UMs to prevent others from obtaining IPR on certain inventions, beyond (or instead of) the UM's purpose of protecting minor and adaptive inventions \citep{heikkila2018role}. Although the motives behind IPR applications are often mixed \citep{granstrand1999economics}, a high proportion of dual applications could be problematic in our context, as we assume that UMs primarily capture imitative and adaptive inventions. We examine the occurrence of dual filings in our data. Identical inventions are grouped at the family level, which allows us to identify dual filings in the data and assess their prevalence. About 0.2\% of Korean domestic UMs are dual-filed as Korean domestic patents, and about 0.3\% of Korean domestic UMs are dual-filed as USPTO patents by Korean entities. These statistics suggest that the proportion of dual filings is negligible in our context.%\footnote{Among the UMs corresponding to dual filings with USPTO patents, 97\% are \emph{directly} linked to the dual-filed USPTO patents in the citation network.} %(i.e., the \textit{distance} equals 1; see Appendix B for the concept of \textit{distance} in our context).}.

%Some of the directly UM-reliant Korean USPTO patents may be such dual filings. Previous research has shown that dual filings with an international patent are rare, but sometimes used by inventors to enhance the scope of IP protection in time and market coverage. Most dual filings are exclusively focused on the domestic market \citep{heikkila2018role}, underpinning the argument that UMs can be useful tools for domestic technological experimentation and incremental innovation tailored to local needs \citep{cahoy2021legal}. 

%\begin{figure*}[htp]
%\centering
%\includegraphics[width=10cm, height=10cm,keepaspectratio]{results/Fig/Fig1.png}
%\caption{Trends of inventive activities}
%  \begin{minipage}{\textwidth}
%    \small\emph{Note.} This Figure presents granted IPRs by the earliest filing year, counted at a family level. \KH{Just a comment: do you think the UM decline in 2007 could be due to the TRIPS agreement, joined by SK in 2007? \url{https://www.wto.org/english/tratop_e/trips_e/amendment_e.htm}; can this have any effect on our results? Maybe it would be interesting to see whether we see the same effect in CN.}
%  \end{minipage}
%\label{FIG:overall_inventive_activities}
%\end{figure*}

\begin{figure*}[htp] \centering
\begin{subfigure}{0.6\textwidth}
\includegraphics[width=10.2cm, height=9.2cm]{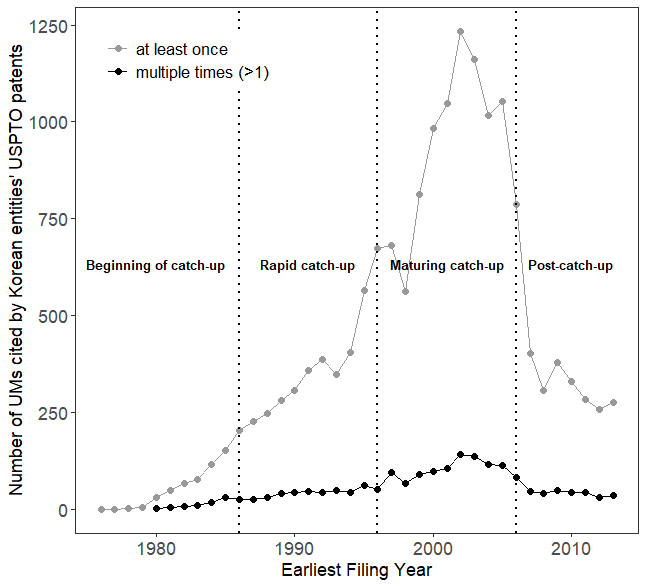}
\caption{\centering UMs cited by the Korean applicants' USPTO patents}\label{Figure5a}
\end{subfigure}
\vskip 1em
\begin{subfigure}{0.8\textwidth}
\centering
\includegraphics[width=10.2cm, height=9.2cm]{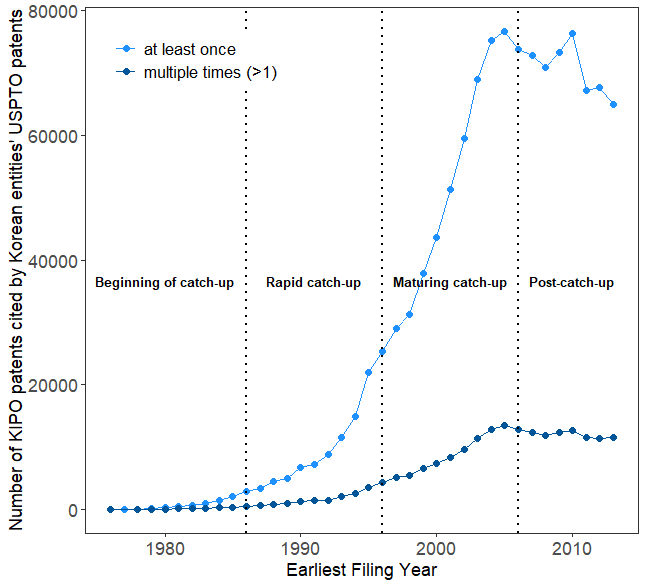}
\captionsetup{justification=centerlast}
\caption{\centering KIPO patents cited by Korean USPTO patents}\label{Figure5b}
\end{subfigure}
\vspace*{2mm}
\captionsetup[subfigure]{font=Large,labelfont=scriptsize}
\caption {Overall trends of citations by Korean USPTO patents}\label{Fig5}
\begin{flushleft}
\begin{minipage}{1\linewidth}
\footnotesize
\emph{Notes. }{The peak of graphs for KIPO patents (Figure \ref{Figure5b}) appears approximately ten years later than that for UMs (Figure \ref{Figure5a}).} 
\end{minipage}
\end{flushleft}
\end{figure*}

Figure \ref{Fig5} shows the evolution of the annual number of UMs and KIPO patents that served as prior art for follow-on frontier technologies (US patents) produced by Korean entities. Figure \ref{Figure5a} shows the number of UMs cited at least once (gray) or multiple times (black) by Korean US patents, indicating that the UMs produced until the mid-2000s increasingly served as an essential basis for the future production of frontier technologies, while their impact has diminished since then. Figure \ref{Figure5b} shows the number of KIPO patents cited at least once (sky blue) or multiple times (dark blue) by the Korean US patents. The peaks of the graphs in this figure appear about ten years later than those in Figure \ref{Figure5a}.

\FloatBarrier
\subsection{Main Results}\label{subsec:main_res}
Table \ref{main} presents the main results of our analyses. Model 0 is a baseline model that includes only control variables. Models 1.1 and 1.2 present the results of testing H1, with Model 1.2 excluding domestic self-citations to better capture the global impact of Korean frontier technologies. Model 2 presents the results of the test of H2.

Model 1.1 shows a positive and significant coefficient for \textit{UM reliance} (0.121$^{***}$). This means that Korean US patents that build on prior knowledge encoded in UMs receive more forward citations than those without any backward link to a UM, supporting our H1. When national self-citations are excluded (Model 1.2), the relevance of UMs decreases slightly but remains positive and statistically significant (0.051$^{***}$). Thus, frontier technologies that rely on UMs are more impactful than those that do not build on UM-protected prior knowledge. The role of \emph{Domestic patent reliance} is insignificant (Model 1.1) or even negative (Model 1.2), indicating that the link between capability building and the domestic patent system is less clear than that with UMs.

In Model 2, we observe a positive and significant moderating role of \textit{Home citation} on the positive effect of UM reliance (0.154$^{***}$). The model shows that Korean US patents tend to receive a higher share of citations from domestic US patents relative to foreign ones when they rely on Korean domestic UMs, supporting our H2. This suggests a disproportionate domestic internalization of positive knowledge externalities arising from UM-reliant frontier technology. This is indicative of unique technological specialization and value internalization, originating from UM-based imitative and adaptive learning experiences. %

\begin{table}[!htbp] \centering 
  \caption{Results of the Cox regression} 
  \label{main}
  \setlength{\tabcolsep}{0.5pt}
  \small
\begin{tabular}{@{\extracolsep{1pt}}p{6.4cm}ccccc} 
\multicolumn{4}{c}{} \\ 
\cline{1-5} 
\\Variable & Model 0 & Model 1.1 & Model 1.2 & Model 2 \\
& Base & H1 & H1 & H2 \\
&  &  & w/o self-citations & \\
\hline \\[-1.8ex]
 \textit{UM reliance} &  & 0.121$^{***}$ & 0.051$^{***}$ & 0.074$^{***}$ \\ 
  &  & (0.008) & (0.010) & (0.010) \\ 
  & & & & \\
 \textit{UM reliance × Home citation} &  &  &  & 0.154$^{***}$ \\ 
  &  &  &  & (0.010) \\ 
  & & & & \\ 
 \textit{Home citation} & $-$0.250$^{***}$ & $-$0.258$^{***}$ &  & $-$0.330$^{***}$ \\ 
  & (0.004) & (0.004) &  & (0.006) \\ 
  & & & & \\ 
 \textit{Domestic patent reliance} & 0.032$^{***}$ & 0.009 & $-$0.020$^{**}$ & 0.012 \\ 
  & (0.009) & (0.009) & (0.010) & (0.009) \\ 
  & & & & \\ 
 \textit{Major applicants} & 0.111$^{***}$ & 0.095$^{***}$ & 0.046$^{***}$ & 0.094$^{***}$ \\ 
  & (0.006) & (0.006) & (0.007) & (0.006) \\ 
  & & & & \\ 
 \textit{Cited citing gap} & $-$0.199$^{***}$ & $-$0.198$^{***}$ & $-$0.261$^{***}$ & $-$0.198$^{***}$ \\ 
  & (0.002) & (0.002) & (0.002) & (0.002) \\ 
  & & & & \\ 
 \textit{Cited citing gap squared} & 0.002$^{***}$ & 0.002$^{***}$ & 0.005$^{***}$ & 0.002$^{***}$ \\ 
  & (0.0001) & (0.0001) & (0.0001) & (0.0001) \\ 
  & & & & \\ 
 \textit{Num prior cit} & 0.003$^{***}$ & 0.003$^{***}$ & 0.002$^{***}$ & 0.003$^{***}$ \\ 
  & (0.0002) & (0.0002) & (0.0002) & (0.0002) \\ 
  & & & & \\ 
 \textit{Num claims} & 0.006$^{***}$ & 0.006$^{***}$ & 0.006$^{***}$ & 0.006$^{***}$ \\ 
  & (0.0004) & (0.0003) & (0.0004) & (0.0003) \\ 
  & & & & \\ 
\textit{Family size} & 0.007$^{***}$ & 0.007$^{***}$ & 0.005$^{***}$ & 0.007$^{***}$ \\ 
  & (0.002) & (0.002) & (0.002) & (0.002) \\ 
  & & & & \\ 
 \textit{Team size} & 0.014$^{***}$ & 0.015$^{***}$ & 0.015$^{***}$ & 0.015$^{***}$ \\ 
  & (0.001) & (0.002) & (0.002) & (0.001) \\ 
  & & & & \\ 
 \textit{Num references} & 0.002$^{***}$ & 0.002$^{***}$ & 0.002$^{***}$ & 0.002$^{***}$ \\ 
  & (0.0002) & (0.0002) & (0.0002) & (0.0002) \\ 
  & & & & \\
  \textit{Field effects} & Yes & Yes & Yes & Yes \\
  \textit{Year effects} & Yes & Yes & Yes & Yes \\ 
\hline \\[-2.0ex]
Events & 2,416,849 & 2,416,849 & 1,748,199 & 2,416,849 \\
Log Likelihood & $-$33,613,471 & $-$33,610,162 & $-$23,628,265 & $-$33,608,779 \\ 
\hline
\end{tabular}\\

    \raggedright
    \footnotesize
    Notes: $^{**}$p$<$0.05; $^{***}$p$<$0.01. Values in parentheses are robust standard errors clustered by focal patents.% Field effects are dummy variables constructed based on the patent CPC section information.\\
\end{table} 
 %updated on 31 Mar 2025
\FloatBarrier
\subsection{Robustness Checks}\label{subsec:robustness_checks}
We conduct a series of robustness checks to address several concerns that could undermine the interpretation of our analyses. 

First, the impact of UMs varies over time, and one could argue that UM-based learning experiences are only relevant during the transition period, without having a lasting impact on shaping a country's technological specialization trajectory. 
Table \ref{main_p06} presents results based on post-2006 USPTO patents to focus on the impact of UMs during the post-catch-up period, as illustrated and discussed in Sections \ref{subsec:case} and \ref{subsec:Results-descriptives}. The post-2006 USPTO patents make up the majority of the patents in our data (200,744 out of 271,515 Korean USPTO patents). The results in Table \ref{main_p06} are consistent with our main results (Table \ref{main}), supporting both hypotheses and confirming the long-term relevance of UM-based learning experiences. 

\begin{table}[!htbp] \centering 
  \caption{Results of the Cox regression (post-2006 USPTO patents only)} 
  \label{main_p06}
  \setlength{\tabcolsep}{0.5pt}
  \small
\begin{tabular}{@{\extracolsep{1pt}}p{6cm}ccccc} 
\multicolumn{4}{c}{} \\ 
\cline{1-5} 
\\Variable & Model 0 & Model 1.1 & Model 1.2 & Model 2 \\
& Base & H1 & H1 & H2 \\
& post-06 & post-06 & post-06 & post-06 \\
&  &  & w/o self-citations & \\
\hline \\[-1.8ex]
 \textit{UM reliance} &  & 0.114$^{***}$ & 0.046$^{***}$ & 0.069$^{***}$ \\ 
  &  & (0.009) & (0.010) & (0.011) \\ 
  & & & & \\ 
 \textit{UM reliance × Home citation} &  &  &  & 0.149$^{***}$ \\ 
  &  &  &  & (0.011) \\ 
  & & & & \\ 
 \textit{Home citation} & $-$0.208$^{***}$ & $-$0.217$^{***}$ &  & $-$0.305$^{***}$ \\ 
  & (0.004) & (0.004) &  & (0.008) \\ 
  & & & & \\  
 \textit{Domestic patent reliance} & 0.066$^{***}$ & 0.020 & $-$0.024 & 0.024 \\ 
  & (0.015) & (0.015) & (0.017) & (0.015) \\ 
  & & & & \\ 
 \textit{Major applicants} & 0.125$^{***}$ & 0.103$^{***}$ & 0.037$^{***}$ & 0.104$^{***}$ \\ 
  & (0.008) & (0.008) & (0.009) & (0.008) \\ 
  & & & & \\
 \textit{Cited citing gap} & $-$0.205$^{***}$ & $-$0.204$^{***}$ & $-$0.446$^{***}$ & $-$0.204$^{***}$ \\ 
  & (0.003) & (0.002) & (0.004) & (0.002) \\ 
  & & & & \\ 
 \textit{Cited citing gap squared} & $-$0.010$^{***}$ & $-$0.010$^{***}$ & 0.010$^{***}$ & $-$0.010$^{***}$ \\ 
  & (0.0003) & (0.0003) & (0.0002) & (0.0003) \\ 
  & & & & \\ 
 \textit{Num prior cit} & 0.003$^{***}$ & 0.003$^{***}$ & 0.002$^{***}$ & 0.003$^{***}$ \\ 
  & (0.0002) & (0.0002) & (0.0002) & (0.0002) \\ 
  & & & & \\ 
 \textit{Num claims} & 0.008$^{***}$ & 0.008$^{***}$ & 0.009$^{***}$ & 0.008$^{***}$ \\ 
  & (0.0004) & (0.0004) & (0.0004) & (0.0004) \\ 
  & & & & \\ 
 \textit{Family size} & 0.006$^{***}$ & 0.006$^{***}$ & 0.004$^{**}$ & 0.006$^{***}$ \\ 
  & (0.002) & (0.002) & (0.002) & (0.002) \\ 
  & & & & \\ 
 \textit{Team size} & 0.010$^{***}$ & 0.010$^{***}$ & 0.011$^{***}$ & 0.010$^{***}$ \\ 
  & (0.002) & (0.002) & (0.002) & (0.002) \\ 
  & & & & \\ 
 \textit{Num references} & 0.001$^{***}$ & 0.001$^{***}$ & 0.001$^{***}$ & 0.001$^{***}$ \\ 
  & (0.0002) & (0.0002) & (0.0002) & (0.0002) \\ 
  & & & & \\
  \textit{Field effects} & Yes & Yes & Yes & Yes \\
  \textit{Year effects} & Yes & Yes & Yes & Yes \\ 
\hline \\[-2.0ex]
Events & 1,433,231 & 1,433,231 & 986,609 & 1,433,231 \\
Log Likelihood & $-$19,177,306 & $-$19,175,355 & $-$12,748,973 & $-$19,174,543 \\ 
\hline
\end{tabular}\\
    \medskip
    \raggedright
    \footnotesize
    Notes: $^{**}$p$<$0.05; $^{***}$p$<$0.01. Values in parentheses are robust standard errors clustered by focal patents.\\ %Field effects are dummy variables constructed based on the patent CPC section information.\\
\end{table} 

 %updated on 31 Mar 2025

Second, to ensure robustness to alternative model setups, we test H1 using a count model based on the five-year forward citation frequency of each patent \emph{i}. Count models are also widely used in innovation studies that use forward citations as the dependent variable (e.g., \citealt{petruzzelli2015determinants}).\footnote{H2 cannot be tested using a count model because the type of forward citation event (domestic vs. foreign) needs to be considered in order to examine H2.} Using negative binomial regressions, we find that H1 is robust to alternative model setups, as shown in Table \ref{rob_count}, with results for Model 1.1 (all patents), Model 1.2 (no self-citations), Model 1.3 (post-2006 patents), and Model 1.4 (post-2006 patents and no self-citations).

%Third, we test the robustness to variations in compiling our binary measure of UM reliance and run a model that includes \textit{UM reliance} as a categorical variable having three categories, direct UM reliance, indirect UM reliance, and no UM reliance (reference group).\footnote{See Appendix B for details of the citation network construction. Direct reliance corresponds to a network \emph{distance}$=1$, and indirect reliance corresponds to a network \emph{distance}$\geq2$.} 
Third, we test the robustness to variations in compiling our binary measure of UM reliance and run a model that includes \textit{UM reliance} as a categorical variable with three categories: direct UM reliance, indirect UM reliance, and no UM reliance (reference group).\footnote{See Appendix B for details on citation network construction. Direct reliance corresponds to a network \emph{distance}$=1$, and indirect reliance corresponds to a network \emph{distance}$\geq2$.} 
This allows us to distinguish between the effects of direct and indirect UM reliance. Table \ref{rob_cate} shows the results for Model 1.1 (all patents), Model 1.2 (post-2006 patents), Model 1.3 (no self-citations), and Model 1.4 (post-2006 patents and no self-citations). The results in Models 1.1 and 1.2 show that indirect reliance plays a larger role than direct reliance, with direct UM reliance being significant only in the post-2006 sample. 
When self-citations are excluded (Models 1.3 and 1.4), indirect UM reliance remains positively significant with a smaller coefficient, while direct UM reliance shows no effect. These results suggest that patents with direct UM reliance contribute mainly to domestic follow-on innovation in the post-catch-up period, while patents with indirect UM reliance have not only a local but also a global impact.

A fourth potential reason for concern is the temporary suspension of UM examination between 1999 and 2006 (see Section \ref{UM:1999-2006}). The suspension may have led to a temporary surge in low-quality UMs granted each year, which could have affected our results. However, the trends in direct and indirect UM reliance appear to be unaffected by the 1999 UM reform (Figure \ref{FIG4}). The annual number of directly UM-reliant patents is roughly constant, and the number of indirectly UM-reliant patents increases almost monotonically. This suggests that the reform probably had no impact, possibly because low-quality UMs are rarely cited and therefore rarely enter our analyses, mitigating concerns about the impact of the reform.

\FloatBarrier

\section{Discussion}\label{sec:Discussion}
This study provides novel insights into the role of IPR systems in catching-up economies, focusing on the UM as a bridging mechanism that supports the transition from the catch-up to the post-catch-up phase. To our knowledge, this is one of the first empirical studies to examine the \emph{long-term} impact of second-tier patent systems in shaping the industrial and technological basis of a country.

We explain how the Korean UM system interacted with the country's catch-up strategy, which prioritized imitative and adaptive learning, specialization, and subsequent diversification in short TCT sectors. Our results show that UM system-based learning in short TCT sectors formed the basis of the country's capabilities to later produce high-impact global frontier technologies in the post-catch-up period. Moreover, the capabilities associated with UM-based learning have become a key source of knowledge localization (i.e., indigenous innovation capabilities) and specialization (i.e., uniqueness of innovation capabilities). Both localization and specialization are hallmarks of advanced economies and key sources of innovation-driven endogenous growth, value-added creation and capture.

In this section, we discuss our contributions to the literature on technological catch-up and IPR regimes (Section \ref{Dis:Evol}) and the implications for innovation policy regarding the conditions under which the UM system can serve as part of a catch-up strategy in low- and middle-income countries (Section \ref{Dis:Indus}).

\subsection{Contributions to the Technological Catch-Up and IPR Regime Literature}
\label{Dis:Evol}
From an evolutionary perspective, technological innovation is often described as a process of technological \textit{variation}, \textit{selection}, and \textit{retention} \citep{nelson1985evolutionary}. This is a process in which innovators, inspired by existing knowledge bases, including science, create various inventions, and only a few of them are selected and succeed in the market. In contrast, in technological catch-up contexts, the sequence is known to be more or less reversed. Innovation often begins with the \textit{selection} of target technologies, followed by \textit{imitative and adaptive learning} efforts to master the target capabilities, and then, if successful, reaches \textit{diversification} in specialized areas \citep{lee2019art}.

This study extends our understanding of technological learning and upgrading during the catch-up period by showing how a second-tier patent system can support \emph{active learning}, with long-lasting effects on a country's indigenous innovation capacity and its technological specialization in the post-catch-up phase \citep{viotti2002national}. Previous studies on catch-up and IPRs have found that the UM system is more appropriate than patents when a country's innovation capacity is weak \citep{ kim2012appropriate,suthersanen2019utility,kang2020intellectual,heikkila2023key}. 
However, given the variation in resource endowments and institutional conditions across developing countries, the suitability of specific catch-up strategies differs accordingly \citep{lee2009both}. The conditions under which the UM regime constitutes an effective mechanism for supporting learning in catch-up economies remain relatively unclear.

Our findings deepen the understanding of the role of different types of IPRs at different stages of economic development by theorizing the mechanisms through which the UM regime can effectively interact with a catching-up country's detour strategy described in previous research \citep{lee2019art, lee2021catching, lee2024economics}. The UM system's short-term of protection and limited scope of protectable subject matter, which excludes process, advanced materials, and biotechnology-related inventions, are consistent with a catching-up strategy that prioritizes specialization in short TCT sectors. Moreover, the UM system's less stringent criteria for obtaining protection compared to the patent system are appropriate to facilitate active learning through imitation and adaptation in short TCT areas. We show how the interaction between such a learning strategy and the UM system can help develop a unique capability base for a country to innovate indigenously at the global frontier, achieve knowledge localization and value internalization \citep{fu2011role, hidalgo2007product, jaffe1993geographic}.

From another perspective, however, our results also suggest that the engine of a post-catch-up economy's production of high-impact frontier technologies tends to remain concentrated in short TCT areas--that is, the sectors shaped by imitative and adaptive learning under the UM regime--despite efforts to develop capabilities in long TCT areas and some progress in this area since reaching the mature catch-up stage (since the early 2000s in the case of Korea) \citep{lee2010ipr}. Since short TCT sectors, by definition, provide niches that are relatively easy for latecomers to enter, the benefits of mature capabilities to produce globally impactful technologies and knowledge localization could quickly erode once latecomers overtake \citep{lee2019art, chang2024dynamics}. In this sense, our results may also indicate possible downsides associated with an active learning history based on the UM system, as it could increase the possibility of lock-in to short TCT industries in the post-catch-up period, exposing the country to the risk of being caught up by the next latecomer.%\footnote{\KH{TBD: ensure consistency with above, Korea made some progress towards long TCT diversification as discussed in other papers. SHould be mentioned here.} --> SJ: Added a reference \citep{lee2010ipr} and a brief explanation: 'some progress'}

%This study presents a case where imitative and adaptive learning, recorded as UMs in targeted technologies, forms the basis of key skills that later inspire novel and impactful innovation at the global frontier. The Korean catch-up case demonstrates a sequence of technological evolution at the country level that differs from the conventional evolutionary view, which describes innovation as an open process of technological variation, selection, and retention \citep{nelson1985evolutionary}. We show how, in catch-up and post-catch-up contexts, the evolution begins with the \textit{selection of target} technologies, followed by \textit{imitative and adaptive learning} efforts to master the target skills, and then, if successful, reaches \textit{specialized variations} at the frontier. This may also have implications for industrial policy in technologically advanced economies, aimed at acquiring and preserving domestic capacities in strategic industries. 

\subsection{Policy Implications}\label{Dis:Indus}
%Our results imply that the traditional rationale for UM as a legal system for encouraging incremental innovation might be too short-term oriented. We demonstrate the \emph{long-term} benefits of UMs as a learning device for imitative and adaptive innovation, especially under a short TCT-focused catch-up strategy. 

Many developing countries, especially those following the Anglo-Saxon common law tradition, do not have a UM system.\footnote{There are a few exceptions. For example, Bangladesh has adopted the UM system in the recently enacted Bangladesh Patent Act 2022.} Given the costs of adopting new IP laws and the lack of a clear rationale, UMs have rarely been considered a development policy option compared to other interventions. Policy discussions on the interactions between IPRs and development have often been limited to the role of IPRs in passive learning through foreign technology transfer, where IPRs were often perceived as barriers and costs to technology access or as irrelevant \citep{falvey2006intellectual, chang2001intellectual, castaldi2024intellectual}.

Our study calls for a more nuanced view on IPRs, identifying a supportive role of second-tier IPRs in a country's active learning process that builds on creatively using and improving imported technology \citep{viotti2002national}. 
The traditional rationale for UMs as legal mechanisms to encourage incremental innovation by inventors with limited capacity appears to be too short-term oriented, overlooking the long-term impact of UMs as part of a national learning strategy. Our findings suggest a direction in which the UM regime can effectively support the long-term process of building globally competitive and local value-added capabilities. Going beyond the existing understanding that UMs protect minor inventions, development and industrial policies could pay more attention to the potential long-term synergies between UM systems and specific types of technological catch-up strategies.

The suitability of specific catch-up strategies varies across contexts. 
Catch-up strategies that take a detour by prioritizing short TCT sectors are particularly relevant for countries with limited natural resources, such as Korea. For countries with sufficient natural resources, such as those in Latin America and Africa, there may be other feasible catch-up paths through high value-added and resource-based industries \citep{lee2021catching, lebdioui2021local}. In this case, UM-based learning support may be less relevant than in catch-up contexts that prioritize short TCT industries. Depending on the portfolio of technological specialization in a country's catch-up strategy, the extent to which the UM regime can contribute to capability accumulation may differ.

The success of technological catch-up is not monocausal, and many factors beyond the detour strategy examined in this study have contributed to the effectiveness of the Korean UM system. Korea's industrial and IPR policies can be described as a well-calibrated mix of export-oriented and protectionist elements, and the UM system was part of this. On the one hand, Korea joined major international IPR and trade agreements early on, which is associated with relatively efficient and standardized administrative and legislative practices \citep{chen2024worldwide}. This created a favorable business environment for foreign investments, which can be essential for gaining access to foreign knowledge in the early stages of economic development. On the other hand, Korean industrial policy included protectionist elements, such as incentive schemes for domestic sourcing of technological components \citep{kim1997imitation}. In this environment, the UM system could function as an enabling mechanism, providing IPR protection for domestic production of high-tech products in short TCT sectors, but not in conflict with international IPR agreements.

Therefore, policymakers in low- and middle-income countries can consider whether the introduction of UM systems can create positive synergies with the countries' overall catch-up strategies and other elements of domestic industrial policy. For countries with an existing UM system that have not been able to reap its benefits, it may be worth strengthening the alignment between the UM system and other aspects of the NIS, including the selection and support of target technology areas, the market environment, and linkages to public research institutes and education.

\section{Future Research and Concluding Remarks}
\label{sec:CM}
Our work has limitations, which point to avenues for future research. In particular, future research may help to understand the role of UMs in different empirical contexts. For example, SMEs have been largely ignored in Korean policy but have played a critical role in Taiwan's catch-up. Although Taiwan also focused on short TCT sectors early, it relied on a network of relatively small firms instead of large firms during its catch-up process \citep{park2006linking}. Many Taiwanese SMEs innovating at the frontier are specialized in relatively less cumulative technologies compared to firms in Korea. Taiwan has also offered UMs as an IP protection mechanism. Investigating the role of the UM system in alternative market environments would be a valuable endeavor for future research. This includes non-catch-up economies, such as various European countries with a strong SME culture and UM systems \citep{heikkila2018empirical}. 

The role of UMs in supporting SMEs or strategic industries in advanced economies is poorly understood. The case for UM seems strong for developing countries with ineffective patent systems, but their value as a `complement' to a well-functioning patent system in an advanced economy remains controversial. The lack of supporting evidence and concerns about legal uncertainty, increased litigation risks, strategic abuse, and regulatory costs led to the abolition of UMs in several developed countries, including the Netherlands, Australia, and Belgium \citep{brack2009utility, johnson2015economic, heikkila2018empirical, suthersanen2019utility, heikkila2023key}. 
Recent studies of UM use in advanced economies suggest that UMs can play an important role in firms' domestic IPR strategies and technological experimentation, providing a rationale for domestically tailored IPR options beyond internationally harmonized patents \citep{cahoy2021legal}.

We do not conclude that the adoption of UMs is a silver bullet, but we show that, under certain conditions, UMs can have a positive impact. In the Korean context analyzed in this study, we show that UMs have left a visible mark on the country's technological capability today.

%Korea's diversification of learning efforts across different product areas may have been important in enabling the production of specialized complex products across various frontiers \citep[][]{hidalgo2007product}. Korea targeted diverse products in its catch-up efforts, including electronic devices, semiconductors, displays, automobiles, oil refineries, shipbuilding, nuclear energy, and machinery.\footnote{In 2021, Korea ranked 3rd by the economic complexity index \citep{hausmann2014atlas} behind Japan and Switzerland. See: \url{https://atlas.cid.harvard.edu/rankings} [accessed on 18/04/2024]} The chaebol conglomerates managed diverse product portfolios and engaged in learning to acquire relevant skills. It is left to future research to examine the role of diversity of target selection when building advanced technological capabilities that span a range of knowledge sources. Such capabilities are hard to acquire for catching-up economies \citep{rosiello2021dynamic}.   

%\section*{Acknowledgements}
%This article benefitted from extensive comments by Jussi Heikkila. The authors further thankfully acknowledge helpful feedback received by staff members of the Gesellschaft f\"ur Internationale Zusammenarbeit (GIZ) and academic peers from the INET complexity group. 

\clearpage
\bibliographystyle{elsarticle-harv}
\bibliography{biblio}

\newpage

\appendix
\FloatBarrier

\renewcommand{\thesection}{\Alph{section}} \setcounter{section}{0}
\renewcommand{\thefigure}{\Alph{section}.\arabic{figure}} \setcounter{figure}{0}
\renewcommand{\thetable}{\Alph{section}.\arabic{table}} \setcounter{table}{0}
\renewcommand{\theequation}{\Alph{section}.\arabic{table}} \setcounter{equation}{0}

\section{Appendix A}
\begin{table}[!htbp] \centering 
   \caption{Top 10 Korean entities patenting in the US}
   \label{TableA1}
   \setlength{\tabcolsep}{0.5pt}
   \small
\begin{tabular}{rllr}
  \hline
 & & Organization & Number of US patents\\ 
  \hline
1 & & SAMSUNG ELECT CO LTD & 120,710 \\ 
  2 & & LG ELECT INC & 36,030 \\ 
  3 & & SAMSUNG DISPLAY CO LTD & 20,005 \\ 
  4 & & HYUNDAI MOTOR CO & 13,542 \\ 
  5 & & LG DISPLAY CO LTD & 10,131 \\ 
  6 & & ELECT \& TELECOM RES INST & 9,911 \\ 
  7 & & SK HYNIX INC & 8,728 \\ 
  8 & & LG CHEM LTD & 7,740 \\ 
  9 & & KIA MOTORS CORP & 7,037 \\ 
  10 & & LG INNOTEK CO LTD & 6,627 \\
  & & Total of the top 10 & 240,461\\
  & & Total & 271,515\\
   \hline
\end{tabular}\\
    \medskip
    \footnotesize
    \textit{Note.} Granted patents counted at the family-level (1976 - 2022, earliest filing date)\\
\end{table}
\label{top10}

\begin{figure*}[htp]
\begin{subfigure}{0.4\textwidth}
\includegraphics[width=8cm, height=7cm,keepaspectratio]{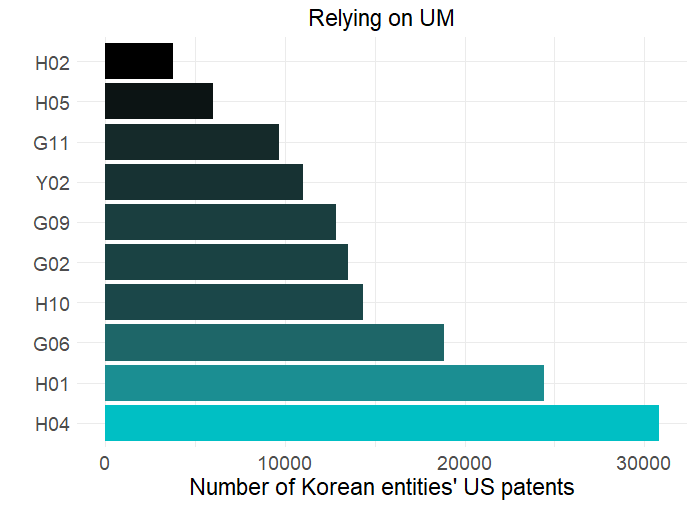}
%\caption{cccc}\label{Figure3a}
\end{subfigure}%
\medskip
\begin{subfigure}{0.4\textwidth}
\centering
\includegraphics[width=8cm, height=7cm,keepaspectratio]{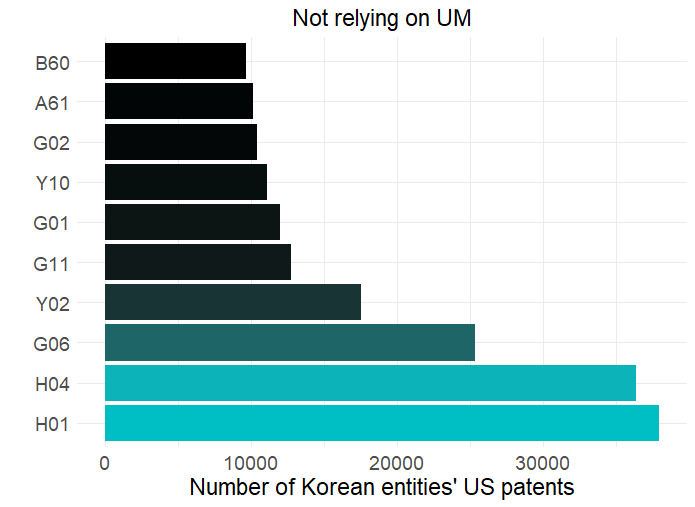}
%\caption{cccc}\label{Figure3}
\end{subfigure}
\vspace*{1mm}
\captionsetup[subfigure]{font=Large,labelfont=scriptsize}
\caption{Major technology fields where Korean entities create US patents}\label{FIGA1}
\begin{flushleft}
\begin{minipage}{1\linewidth}
\footnotesize
\emph{Note 1.} The figures display the top 10 fields (CPC classes) in which Korean entities create US patents. The left and right figures represent patents that rely on the UM and those that do not rely on the UM, respectively. Six out of the ten fields (H04, H01, G06, Y02, G11, and G02) overlap between the two groups.\\
\emph{Note 2.} CPC class definition: H04 Electric Communication Technique; H01 Electric Elements; G06 Computing; H10 Semiconductor Devices; G02 Optics; G09 Education, Cryptography, Display, Advertising, Seals; Y02 Technologies for Mitigation or Adaptation Against Climate Change; G11 Information Storage; H05 Electric Techniques; H02 Generation, Conversion or Distribution of Electric Power; G01 Measuring, Testing; Y10 Technical Subjects Covered by Former USPC; A61 Medical of Veterinary Science, Hygiene; B60 Vehicles in General
\end{minipage}
\end{flushleft}
\end{figure*}

\begin{table}[ht]\centering 
  \caption{Correlation Matrix}
  \label{cor}
\adjustbox{angle=0,scale=0.7}{
\begin{tabular}{p{7cm}rrrrrrrrrrr}
  \hline
 & Mean & S.D. & \textit{(4)} & \textit{(5)} & \textit{(6)} & \textit{(7)} & \textit{(8)} & \textit{(9)} \\ 
  \hline
\textit{(1) UM reliance (1: Yes, 0: No)} & 0.39 & &  &  & & &  &  & \\ 
  \textit{(2) Domestic patent reliance (1: Yes, 0: No)} & 0.90 & &  &  &  & &  &  & \\ 
  \textit{(3) Major applicants (1: Yes, 0: No)} & 0.75 & &  &  &  & &  &  & \\ 
  \textit{(4) Cited citing gap} & 5.13 & 5.60 & 1.000 & &  & &  &  & \\ 
  \textit{(5) Num prior cit} & 21.53 & 74.12 & 0.111 & 1.000 &  & &  &  & \\ 
  \textit{(6) Num claims} & 16.91 & 10.08 & -0.072 & 0.092 & 1.000 & &  &  & \\ 
  \textit{(7) Family size} & 4.69 & 6.02 & -0.030 & 0.110 & 0.151 & 1.000 &  &  & \\ 
  \textit{(8) Team size} & 3.17 & 2.20 & -0.123 & 0.065 & 0.085 & 0.094 & 1.000 &  & \\ 
  \textit{(9) Num refs} & 20.28 & 28.66 & -0.076 & 0.125 & 0.234 & 0.378 & 0.130 & 1.000 & \\  
   \hline
\end{tabular}
}\\
    \medskip
    \raggedright
    \footnotesize
    Notes: \textit{UM reliance}, \textit{Domestic patent reliance}, and \textit{Major applicants} are binary variables. Pearson correlation coefficients are reported for numerical variables. All coefficients are significant at p$<$0.01 level.\\
\end{table}

\newpage

\section{Appendix B}
\FloatBarrier

To get our independent variable, \textit{UM reliance}, we utilize the concept of distance metric by \citet{ahmadpoor2017dual}, in which authors measure the minimum distance from science to technology in a specific technology field by using a citation network connecting two worlds: research publications (a proxy of science) and patents (technology). 

Drawing from this concept, in our context, we construct a citation network consisting of Korean domestic UMs and Korean entities' US patents. We compute each patent's minimum distance to one of the domestic UMs in the constructed network by considering a direct citation link as a unit distance. If a patent \textit{i} can be directly connected to one of the Korean UMs, the distance, $d_{\text{i}}$, is 1 (direct UM reliance). If a patent \textit{i} can be connected to one of the Korean UMs by passing through at least one prior work, the distance, $d_{\text{i}}$ is 2. If the distance is greater than or equal to 2, it corresponds to indirect UM reliance. In our data, the maximum $d_{\text{i}}$ for indirect UM reliance is 16. If a patent \textit{i} cannot be connected to any Korean domestic UMs at any distance, the distance is, by definition, infinite, and it is considered non-reliance.
%To address the infinite values, we find the maximum distance $d_{\text{max}}$\ in our data and assign the value to patents that are disconnected from the UM. $d_{\text{max}}$\ is 16 in our data. Finally, to get a patent \textit{i}'s reliance on the UM, we assign reverse ordered distance to each patent (i.e., $d_{\text{max}}$+1-$d_{\text{i}}$), making the higher value indicates a higher reliance on the UM system. As our $d_{\text{max}}$\ is 16, \textit{UM reliance$_{cont.}$} used in robustness check (Table \ref{rob_con}) ranges between 1 and 16. 

As for the binary \textit{UM reliance} used in our main analysis (Table \ref{main} and \ref{main_p06}), if a patent \textit{i} is connected to Korean UMs at any distance ($d_{\text{i}} \geq 1$), its value is 1; otherwise, it is 0. The \textit{UM reliance} in Table \ref{rob_cate} distinguishes the three categories: direct UM reliance ($d_{\text{i}} = 1$), indirect UM reliance ($d_{\text{i}} \geq 2$), and non-reliance. 

The control variable \emph{Domestic patent reliance} is computed based on the same logic, but based on a network of citations from Korean US patents to Korean domestic patents (instead of UMs).

\newpage

\section{Appendix B}
\FloatBarrier

\begin{table}[!htbp] \centering 
  \caption{Robustness check using negative binomial models (DV: 5-year forward citation counts)} 
  \label{rob_count}
  \setlength{\tabcolsep}{0.7pt}
  \small
\begin{tabular}{@{\extracolsep{1.5pt}}p{4cm}cccccc} 
\multicolumn{5}{c}{} \\ 
\cline{1-6} 
\\Variable & Model 0 & Model 1.1 & Model 1.2 & Model 1.3 & Model 1.4\\
& Base & H1 & H1 & H1 & H1\\
&  &  &  & post-06 & post-06\\
&  &  & w/o self-citations & & w/o self-citations\\
\hline \\[-1.8ex]
 \textit{UM reliance} &  & 0.243$^{***}$ & 0.093$^{***}$ & 0.274$^{***}$ & 0.125$^{***}$ \\ 
  &  & (0.005) & (0.006) & (0.006) & (0.007) \\ 
  & & & & & \\ 
 \textit{Domestic patent reliance} & 0.107$^{***}$ & 0.061$^{***}$ & $-$0.040$^{***}$ & 0.010 & $-$0.088$^{***}$ \\ 
  & (0.008) & (0.008) & (0.009) & (0.012) & (0.014) \\ 
  & & & & & \\ 
 \textit{Major applicants} & 0.328$^{***}$ & 0.290$^{***}$ & 0.218$^{***}$ & 0.344$^{***}$ & 0.241$^{***}$ \\ 
  & (0.005) & (0.005) & (0.006) & (0.007) & (0.008) \\ 
  & & & & & \\ 
 \textit{Num claims} & 0.021$^{***}$ & 0.021$^{***}$ & 0.021$^{***}$ & 0.025$^{***}$ & 0.025$^{***}$ \\ 
  & (0.0003) & (0.0003) & (0.0003) & (0.0004) & (0.0005) \\ 
  & & & & & \\ 
 \textit{Family size} & 0.050$^{***}$ & 0.046$^{***}$ & 0.039$^{***}$ & 0.051$^{***}$ & 0.044$^{***}$ \\ 
  & (0.001) & (0.001) & (0.001) & (0.001) & (0.001) \\ 
  & & & & & \\ 
 \textit{Team size} & 0.035$^{***}$ & 0.036$^{***}$ & 0.035$^{***}$ & 0.031$^{***}$ & 0.028$^{***}$ \\ 
  & (0.001) & (0.001) & (0.001) & (0.001) & (0.001) \\ 
  & & & & & \\ 
 \textit{Num references} & 0.009$^{***}$ & 0.007$^{***}$ & 0.008$^{***}$ & 0.007$^{***}$ & 0.008$^{***}$ \\ 
  & (0.0001) & (0.0001) & (0.0002) & (0.0002) & (0.0002) \\ 
  & & & & & \\ 
 Constant & $-$0.349 & $-$0.336 & $-$0.304 & 0.388$^{***}$ & 0.348$^{***}$ \\ 
  & (0.987) & (0.986) & (1.065) & (0.021) & (0.024) \\ 
  & & & & & \\
  \textit{Field effects} & Yes & Yes & Yes & Yes & Yes\\
  \textit{Year effects} & Yes & Yes & Yes & Yes & Yes\\ 
\hline \\[-1.8ex] 
Observations & 271,515 & 271,515 & 271,515 & 200,744 & 200,744 \\ 
Log Likelihood & $-$737,400 & $-$736,252 & $-$643,247 & $-$524,180 & $-$451,343 \\ 
\hline 
\hline \\[-1.8ex]
\end{tabular}\\
    \medskip
    \raggedright
    \footnotesize
    Notes: $^{**}$p$<$0.05; $^{***}$p$<$0.01. Time-varying controls in the original model (i.e., \emph{Cited citing gap}, \emph{Cited citing gap squared}, and \emph{Num prior cit}) are inappropriate in the count model setup, so they are excluded.\\% Field effects are dummy variables constructed based on the patent CPC section information.\\
\end{table} 

 %added on 21 Mar 2025

\begin{table}[!htbp] \centering 
  \caption{Results of the Cox regression (distinguishing direct and indirect UM reliance)} 
  \label{rob_cate}
  \setlength{\tabcolsep}{0.5pt}
  \small
\begin{tabular}{@{\extracolsep{1pt}}p{6cm}ccccc} 
\multicolumn{4}{c}{} \\ 
\cline{1-5} 
\\Variable & Model 1.1 & Model 1.2 & Model 1.3 & Model 1.4 \\
& H1 & H1 & H1 & H1 \\
& & post-06 & & post-06 \\
&  &  & w/o self-citations & w/o self-citations\\
\hline \\[-1.8ex]
 \textit{UM reliance (direct)} & 0.018 & 0.048$^{***}$ & $-$0.024 & $-$0.001 \\ 
  & (0.011) & (0.015) & (0.013) & (0.017) \\ 
  & & & & \\ 
 \textit{UM reliance (indirect)} & 0.138$^{***}$ & 0.120$^{***}$ & 0.065$^{***}$ & 0.051$^{***}$ \\ 
  & (0.009) & (0.009) & (0.010) & (0.010) \\ 
  & & & & \\ 
 \textit{Home citation} & $-$0.259$^{***}$ & $-$0.217$^{***}$ &  &  \\ 
  & (0.004) & (0.004) &  &  \\ 
  & & & & \\ 
 \textit{Domestic patent reliance} & 0.008 & 0.018 & $-$0.020$^{**}$ & $-$0.025 \\ 
  & (0.009) & (0.015) & (0.010) & (0.018) \\ 
  & & & & \\ 
 \textit{Major applicants} & 0.091$^{***}$ & 0.100$^{***}$ & 0.042$^{***}$ & 0.035$^{***}$ \\ 
  & (0.006) & (0.008) & (0.007) & (0.010) \\ 
  & & & & \\ 
 \textit{Cited citing gap} & $-$0.198$^{***}$ & $-$0.204$^{***}$ & $-$0.261$^{***}$ & $-$0.446$^{***}$ \\ 
  & (0.002) & (0.002) & (0.002) & (0.004) \\ 
  & & & & \\ 
 \textit{Cited citing gap squared} & 0.002$^{***}$ & $-$0.010$^{***}$ & 0.005$^{***}$ & 0.010$^{***}$ \\ 
  & (0.0001) & (0.0003) & (0.0001) & (0.0002) \\ 
  & & & & \\ 
 \textit{Num prior cit} & 0.003$^{***}$ & 0.003$^{***}$ & 0.002$^{***}$ & 0.002$^{***}$  \\ 
  & (0.0002) & (0.0002) & (0.0002) & (0.0002) \\ 
  & & & & \\ 
 \textit{Num claims} & 0.006$^{***}$ & 0.008$^{***}$ & 0.006$^{***}$ & 0.009$^{***}$ \\ 
  & (0.0003) & (0.0004) & (0.0004) & (0.0004) \\ 
  & & & & \\ 
 \textit{Family size} & 0.007$^{***}$ & 0.006$^{***}$ & 0.005$^{***}$ & 0.004$^{**}$ \\ 
  & (0.002) & (0.002) & (0.002) & (0.002) \\ 
  & & & & \\ 
 \textit{Team size} & 0.014$^{***}$ & 0.010$^{***}$ & 0.015$^{***}$ & 0.011$^{***}$ \\ 
  & (0.002) & (0.002) & (0.002) & (0.002) \\ 
  & & & & \\ 
 \textit{Num references} & 0.002$^{***}$ & 0.001$^{***}$ & 0.002$^{***}$ & 0.001$^{***}$ \\ 
  & (0.0002) & (0.0002) & (0.0002) & (0.0002) \\ 
  & & & & \\
  \textit{Field effects} & Yes & Yes & Yes & Yes \\
  \textit{Year effects} & Yes & Yes & Yes & Yes \\ 
\hline \\[-2.0ex]
Events & 2,416,849 & 1,433,231 & 1,748,199 & 986,609 \\
Log Likelihood & $-$33,609,555 & $-$19,175,241 & $-$23,628,025 & $-$12,748,932 \\ 
\hline
\end{tabular}\\
    \medskip
    \raggedright
    \footnotesize
    Notes: $^{**}$p$<$0.05; $^{***}$p$<$0.01. Values in parentheses are robust standard errors clustered by focal patents.\\ %Field effects are dummy variables constructed based on the patent CPC section information.\\
\end{table} 
 %added on 31 Mar 2025

\end{document}